\documentclass[a4paper,11pt]{article}
\pdfoutput=1 % if your are submitting a pdflatex (i.e. if you have
\usepackage{jcappub} % for details on the use of the package, please
\usepackage{ulem}

\usepackage[T1]{fontenc} % if needed

\usepackage{color}
\definecolor{DarkGreen}{rgb}{.2, .65, .2}

\title{Examining the influence of anisotropy on the fundamental mode of nonradial oscillation in neutron stars on a complete general relativistic scheme}

\author[a,b,1]{Jos\'e D. V. Arba\~nil\note{Corresponding author.}}
\author[c]{Gabriel O. Cavalheiro}
\author[d]{Victor B. T. Alves}
\author[e]{Juan M. Z. Pretel}
\author[d,f]{C\'esar O. V. Flores}
\author[c]{C\'esar H. Lenzi}

\affiliation[a]{Departamento de Ciencias, Universidad Privada del Norte, \\Avenida el Sol 461 San Juan de Lurigancho, 15434 Lima, Peru.}
\affiliation[b]{Facultad de Ciencias F\'isicas, Universidad Nacional Mayor de San Marcos, \\Avenida Venezuela s/n Cercado de Lima, 15081 Lima,  Peru.}
\affiliation[c]{Departamento de F\'isica, Instituto Tecnol\'ogico de Aeron\'autica, \\S\~ao Jos\'e dos Campos, S\~ao Paulo, CEP 12228-900, Brazil.}
\affiliation[d]{Programa de P\'os-gradua\c{c}\~ao em F\'isica - CCET
Universidade Federal do Maranh\~ao,
65080-805, São Lu\'is, Maranh\~ao, Brazil.}
\affiliation[e]{Centro Brasileiro de Pesquisas F{\'i}sicas, \\Rua Dr. Xavier Sigaud, 150 URCA, Rio de Janeiro CEP 22290-180, RJ, Brazil.}
\affiliation[f]{Universidade Estadual da Regi\~ao Tocantina do Maranh\~ao, \\Centro de Ci\^encias Exatas, Naturais e Tecnol\'ogicas,  Imperatriz, Maranh\~ao, CEP 65901-480, Brazil.}

\emailAdd{jose.arbanil@upn.pe}
\emailAdd{gabrieloliveira8551@gmail.com}
\emailAdd{victor.bruno@discente.ufma.br}
\emailAdd{juan04manuel91@gmail.com}
\emailAdd{cesar.vasquez@uemasul.edu.br}
\emailAdd{chlenzi@ita.br}

\abstract{
The anisotropic influence on the $f$-mode frequency of oscillations and dimensionless tidal deformability of neutron stars is analyzed by employing the nonradial oscillation equations for the complete general relativity frame and tidal deformability equations, which are derived and modified from their standard form to introduce the anisotropic factor. The fluid inside the compact star obeys an equation of state constructed by matching microscopic nuclear and perturbative QCD calculations through a piecewise polytropic interpolating scheme. For the anisotropic profile, we use a local anisotropy which is regular along the whole star and vanishes both at the center and on the star's surface. We show that the \(f\)-mode oscillation frequency and dimensionless tidal deformability are noticeably affected by anisotropy. Finally, we investigate the correlation between the dimensionless tidal deformability inferred from the GW$170817$ event and the anisotropy parameter.}

\begin{document}

\maketitle
\flushbottom
%\keyword{Compact star, nonradial oscillations, anisotropy.}

%\maketitle

%%%%%%%%%%%%%%%%%%%%%%%%%%%%%%%%%%%%%%%%%
\section{\label{sec:level1} Introduction}
%%%%%%%%%%%%%%%%%%%%%%%%%%%%%%%%%%%%%%%%%

%Since the first detection of gravitational wave (GW) signals from a binary neutron star (NS) merger, known as GW$170817$ and reported by the LIGO-Virgo Collaboration (LVC) \cite{abbott2017_4}, numerous researchers have made significant efforts to explore the microphysics of NSs in more depth, providing stringent constraints on the equation of state (EOS) of such stars and significantly narrowing the range of possible theoretical models. In other words, the GW analysis provides a promising avenue for probing the behavior of ultra-dense matter under extreme conditions. This can be achieved by studying the EOS through the gravitational waves generated during the merger of two compact stars. 

Since the first detection of gravitational-wave (GW) signals from a binary neutron star (NS) merger, known as GW$170817$ and reported by the LIGO-Virgo Collaboration (LVC) \cite{abbott2017_4}, numerous studies have explored the microphysics of NSs in greater depth, providing increasingly stringent constraints on the equation of state (EOS) of dense matter \cite{Kurkela_2014, Fraga_2014, Bauswein2017, 2018PhRvL.121p1101A, 2018PhRvC..98c5804M}. These investigations have significantly narrowed the range of viable theoretical models.

Nevertheless, the GW$170817$ event is only one example among the different types of compact binary mergers observed with ground-based gravitational-wave detectors, including binary black hole mergers \cite{abbott2016, abbott2016_2, abbott2017, abbott2017_2} and binary NS systems, the latter providing the first multi-messenger observation in the case of GW$170817$ \cite{abbott2017_5}. Furthermore, future space-based GW observatories like DECIGO are expected to open a new window for probing new physics related to highly deformed compact objects, such as NSs, hybrid stars, and magnetars, which are currently inaccessible to ground-based detectors \cite{Miller2025PRD}.

%{\bf Asteroseismology is a powerful theoretical tool for gaining a deeper understanding of the behavior of dense matter inside compact stars \cite{pulsating2, pulsating4, pulsating5, flores2017}. In that regard,} nonradial oscillations have been extensively studied, especially in light of upcoming third-generation {\bf gravitational-wave} detectors, which are expected to be sensitive to such signals, or due to the possibility of observing their signatures through resonances in binary star systems. Numerous studies have explored pulsation modes assuming various compositions of the isotropic fluid within compact stars, including deconfined quark matter \cite{sotani_harada2003}, color-flavor-locked strange quark matter \cite{flores_lugones2017}, hadronic matter \cite{andersson_kokkotas_1998, benhar_ferrari_2004}, scenarios with density discontinuities \cite{sotani2001}, and hadronic matter admixed with dark matter \cite{flores_lenzi_dutra_2004}, among others. These studies consistently show that nonradial oscillation modes are highly sensitive to the microphysical characteristics of the fluid within compact stars, leading to significant variations in their behavior. 

If nonradial oscillations of compact stars were to be detected, then gravitational wave asteroseismology would be a powerful theoretical tool for gaining a deeper understanding of the behavior of dense matter inside compact stars \cite{pulsating2, pulsating4, pulsating5, flores2017}. Numerous studies have explored nonradial oscillation modes assuming various compositions of the isotropic fluid within compact stars, including deconfined quark matter \cite{sotani_harada2003}, color-flavor-locked strange quark matter \cite{flores_lugones2017}, hadronic matter \cite{andersson_kokkotas_1998, benhar_ferrari_2004}, scenarios with density discontinuities \cite{sotani2001}, and hadronic matter admixed with dark matter \cite{flores_lenzi_dutra_2004}, among others. These studies consistently show that nonradial oscillation modes are highly sensitive to the microphysical characteristics of the fluid within compact stars, leading to significant variations in their behavior. {It is also worth mentioning that third-generation instruments are expected to be more sensitive to continuous GW signals from, e.g., $r$-mode or $f$-mode instabilities, from NS mountains \cite{Gittins2020, Gittins2021} or due to resonances in compact binary systems \cite{Pnigouras2025}.}

Nonetheless, theoretical evidence suggests that various phenomena may induce anisotropy in fluids under extreme conditions \cite{paper3, CARTER1998478, Yazadjiev2012}; see also the review articles \cite{HERRERA199753, KUMAR2022101662} for extensive discussions on anisotropic pressure in self-gravitating systems. For instance, the geometry of $\pi^-$ modes indicates that anisotropic pressure distributions may arise in NSs as a manifestation of pion-condensed phase configurations \cite{sawyer1973}. Additionally, anisotropy can be generated by the presence of solid or superfluid cores within NSs \cite{ruderman1972,heiselberg2000}. Furthermore,  anisotropic stresses naturally emerge in self-gravitating systems composed of complex scalar fields, such as boson stars \cite{gleiser1988,gleiser1989}. In addition, it has been shown in recent years that the inclusion of anisotropy in relativistic star systems leads to theoretical predictions that are consistent with several observational mass-radius measurements and tidal deformability constraints \cite{curi2022non, Das2023MNRAS, PRETEL2024138375, Becerra2024}. In particular, the impact of pressure anisotropy on the tidal deformability of compact stars has been previously studied by Biswas and Bose \cite{Biswas2019}, who showed that anisotropy can significantly modify the tidal response while remaining compatible with GW$170817$ constraints.

Motivated by these studies, in the present work, we investigate the impact of anisotropy on the $f$-mode (fundamental mode) oscillation frequency and tidal deformability of NSs. For the anisotropic profile, we consider the relation $\sigma=\alpha p_r\mu^{a}$, where $p_r$ denotes the radial pressure and $\mu=2m/r$ is a compactness-related quantity, with $a=2$. For the EOS, we employ the microscopic nuclear calculations with perturbative Quantum Chromodynamics (pQCD) results, connected through a piecewise polytropic interpolation framework \cite{Kurkela_2014,PhysRevD.81.105021,Fraga_2014}. To this end, we derive the nonradial oscillation equations and the regularity conditions for the perturbation variables near the center within the framework of general relativity for anisotropic matter. The nonradial oscillation equations derived in this study are consistent with those reported in \cite{2024PhRvD.110h3020L,2026PhRvD.113d3025G,2026arXiv260106962Y}, but differ from the expressions presented in \cite{modal_2024, mondal2025}. Such equations, together with the deformability equations, are then integrated to assess the influence of anisotropy on the fundamental oscillation modes and stellar tidal deformability.

It is worth noting that, in a similar context, the impact of the same anisotropy profile with $a=2$ on the quasinormal modes of NSs has been investigated in \cite{2024PhRvD.110h3020L}, using the polytropic EOS II \cite{doneva2012} and SLy4 \cite{douchin2001}, as well as in \cite{2026arXiv260106962Y}, where the WFF1 \cite{wiringa1988}, SLy4, and MS1 \cite{muller1996} EOSs were employed. Similarly, the anisotropic pressure effects on the fundamental oscillation modes of NSs have been studied with $a=2$ in \cite{2026PhRvD.113d3025G}, considering the WFF1, MS1, and MPA1 \cite{muther1987} EOSs, and with $a=1$ in \cite{modal_2024}, using the BSk19 \cite{goriely2010} and BSk21 \cite{pearson2012} EOSs. In addition, anisotropy effects have been explored in quark stars \cite{mondal2025}, both for the MIT bag model EOS describing non-interacting quark matter and for EOSs that include inter-quark interactions.

%The nonradial oscillation equations derived in this study are consistent with those reported in \cite{2024PhRvD.110h3020L,2026PhRvD.113d3025G,2026arXiv260106962Y}, but differ from the expressions presented in \cite{modal_2024, mondal2025}. The difference with the last two studies occurs because, in the anisotropic case, the first law of thermodynamics needs to consider that the work done on the fluid varies with direction. 

%{\bf A key property in the study of compact objects is their tidal deformability (see, e.g., \cite{flanagan,hinderer_2008,damour2009,postnikov2010,poisson}). This quantity encodes valuable information about the equation of state of dense matter, as reflected in the gravitational-wave signals emitted by these stars. Importantly, by examining the dimensionless tidal deformability, current observations-particularly from GW$170817$-enable us to place meaningful constraints on theoretical models.}

This manuscript is structured as follows: Section \ref{section} presents the hydrostatic equilibrium equations and the nonradial oscillation equations, including the anisotropic factor. Section \ref{section3} shows the EOS and the anisotropic profile equation. In Section \ref{sectio_numerical_method}, we describe the numerical method employed, and in Section \ref{results}, we report the results in the analysis of anisotropy effects on the equilibrium, oscillation frequencies, and tidal deformability. In Section \ref{conclusions}, we conclude. {Finally, in Appendix \ref{Appendix_A}, the perturbation variables and the general form of the linearized perturbation equations for the anisotropic case are presented. In Appendix \ref{Appendix_B}, the perturbation functions explicitly expanded at the star's center are presented, which are important to guarantee finite and numerically stable solutions.} Throughout this article, we adopt geometric units by setting $G=c=1$ to simplify the equations and facilitate numerical calculations.

%%%%%%%%%%%%%%%%%%%%%%%%%%%%%%%%%%%%%%%%%%%%%%%%%%%%%%%%%%%%%
\section{General relativistic equations}\label{section}
%%%%%%%%%%%%%%%%%%%%%%%%%%%%%%%%%%%%%%%%%%%%%%%%%%%%%%%%%%%%%
\subsection{The static equilibrium equations}\label{subsection1}

For completeness, we start by writing the Einstein field equation 
\begin{equation}\label{einstein_eq}
    G_{\mu\nu}=8\pi T_{\mu\nu},
\end{equation}
where the Greek indices $\mu,\nu,$ etc., run from $0$ to $3$. $G_{\mu\nu}$ represents the Einstein tensor and $T_{\mu\nu}$ stands for the energy-momentum tensor of an anisotropic fluid, given by \cite{arbanil_malheiro_2016}:
\begin{equation}\label{tem_background}
T_{\mu\nu}=\left(p_r+\sigma\right)g_{\mu\nu}+(\rho+p_r+\sigma)u_{\mu}u_{\nu}-\sigma k_{\mu}k_{\nu}.
\end{equation}
The variables $p_r$, $\rho$, and $\sigma$ depict, respectively, the radial pressure, the energy density, and the anisotropic factor. $g_{\mu\nu}$, $u_{\mu}$, and $k_{\mu}$ represent the spacetime metric tensor, the fluid's four-velocity, and the radial unit vector, respectively. In addition, the four-velocity and the radial unit vector follow the equalities:
\begin{equation}
u_{\mu}u^{\mu}=-1,\quad k_{\mu}k^{\mu}=1,\quad {\rm and}\quad k_{\mu}u^{\mu}=0.
\end{equation}

To analyze the static equilibrium configuration of spherically symmetric static compact stars, we consider the line element in the Schwarzschild-like coordinates ($t, r,\theta,\phi$) as follows:
\begin{equation}\label{line_element_background}
ds^2=-e^{2\Psi}dt^2+e^{2\Lambda}dr^2+r^{2}\left(d\theta^2+\sin^2\theta d\phi^2\right),
\end{equation}
where $\Psi=\Psi(r)$ and $\Lambda=\Lambda(r)$ are functions of the radial coordinate $r$ only.

Substituting the energy-momentum tensor \eqref{tem_background} and the metric \eqref{line_element_background} into the Einstein field equation, we derive the hydrostatic stellar structure equations:
\begin{eqnarray}
&&\frac{dp_r}{dr}=-\frac{\rho m}{r^2}\left[1+\frac{p_r}{\rho}\right]\left[1+\frac{4\pi r^3 p_r}{m}\right]e^{2\Lambda}+\frac{2\sigma}{r},\label{tov}\\
&&\frac{dm}{dr}=4\pi r^2\rho,\label{conservatio_mass}\\
&&\frac{d\Psi}{dr}=-\frac{1}{p_r+\rho}\frac{dp_r}{dr}+\frac{2\sigma}{r(p_r+\rho)}\label{phi_eq},
\end{eqnarray}
with

\begin{equation}\label{eq_phi}
e^{2\Lambda}=\left(1-\frac{2m}{r}\right)^{-1}.
\end{equation}

As usual, the parameter $m$ represents the mass within the sphere of radius $r$. Eq.~\eqref{tov} is the hydrostatic equilibrium equation, also known as the Tolman-Oppenheimer-Volkoff equation \cite{tolman,oppievolkoff}, 
modified from its standard form to include the anisotropic factor $\sigma$, see Ref.~\cite{bowers1974}. 

Eqs.~\eqref{tov}-\eqref{phi_eq} are integrated from the center ($r=0$) to the surface of the star ($r=R$). This solution starts at the center of the star $r=0$, where:
\begin{eqnarray}
&&m(0)=0,\quad\rho(0)=\rho_c,\quad p_r(0)=p_{rc},\quad\sigma(0)=0,\quad {\rm and}\quad\Psi(0)=\Psi_c,
\end{eqnarray}
and the star's surface is determined by $p_r(r=R)=0$. Moreover, the interior line element connects smoothly with the exterior Schwarzschild vacuum metric at the star's surface. Thus, the inner and outer potential metric functions are connected by the relation:

\begin{equation}\label{eq_phi_s}
e^{2\Psi(R)}=e^{-2\Lambda(R)}=1-\frac{2M}{R},
\end{equation}
with $M$ representing the total mass of the star. Eq.~\eqref{eq_phi_s} depicts the boundary condition of $\Psi(R)$ at the star's surface.

\subsection{Nonradial perturbation equations}\label{subsection2}

The study of the nonradial oscillations is investigated by decomposing the perturbed metric into a background metric $g^{(0)}_{\mu\nu}$, whose components are found in Eq.~\eqref{line_element_background}, plus the metric perturbation $h_{\mu\nu}$. This decomposition can be placed into the form:
\begin{equation}\label{background_perturbed_metric}
    g_{\mu\nu}=g^{(0)}_{\mu\nu}+h_{\mu\nu}.
\end{equation}

In pulsating compact stellar configurations, the fluid and spacetime dynamics are described by the perturbed Einstein field equation and conservation of the energy-momentum tensor
\begin{align}
    \delta G^{\mu}_{\varphi} = 8\pi \delta T^{\mu}_{\varphi}, \label{perturbed_FE} \\
    \delta\left(\nabla_{\mu}T^{\mu}_{\varphi}\right) = 0,
\end{align}
with
\begin{align}
\delta G^{\mu}_{\varphi}=&\ g^{(0)\mu\beta}\delta R_{\varphi\beta}-\frac{1}{2}\delta_{\varphi}^{\mu}\left(g^{(0)\alpha\beta}\delta R_{\alpha\beta}-h^{\alpha\beta}R^{(0)}_{\alpha\beta}\right)-\frac{1}{2}g^{(0)\mu\beta}h_{\varphi\beta}R^{(0)}-h^{\mu\beta}G^{(0)}_{\varphi\beta},\label{delta_G}\\
\delta T^{\mu}_{\varphi}=&\ \left(\delta\rho+\delta p_r+\delta\sigma\right)u^{\mu}u_{\varphi}+\left(\rho+p_r+\sigma\right)\delta u^{\mu}u_{\varphi}+\left(\rho+p_r+\sigma\right)u^{\mu}\delta u_{\varphi}+\delta p_r g^{(0)}_{\nu\varphi}g^{(0)\mu\nu}\nonumber\\
&+\delta\sigma g^{(0)}_{\nu\varphi}g^{(0)\mu\nu}+(p_r+\sigma)h_{\nu\varphi}g^{(0)\mu\nu}-(p_r+\sigma)g^{(0)}_{\nu\varphi}h^{\mu\nu}-\delta\sigma g^{(0)}_{\nu\varphi}k^{\mu}k^{\nu}-\sigma h_{\varphi\nu}k^{\mu}k^{\nu}\nonumber\\
&-\sigma g^{(0)}_{\varphi\nu}\delta k^{\mu}k^{\nu}-\sigma g^{(0)}_{\varphi\nu}k^{\mu}\delta k^{\nu}.\label{delta_T}
\end{align}

%To investigate the effects of anisotropy on the fluid pulsation mode emitted by compact stars, the perturbed line element is placed into the form:
%
%\begin{equation}\label{complete_line_element}
%ds^2=-\left(e^{2\Psi}-%h_{00}\right)dt^2+2h_{10}dtdr+\left(e^{2\Lambda}+h_{11}\right)dr^2+\left(r^{2}+h_{22}\right)\left(d\theta^2+\sin^2\theta d\phi^2\right),
%\end{equation}
%where $\Psi=\Psi(r)$ and $\Lambda=\Lambda(r)$ are the background metric functions and $h_{00}=h_{00}(t,r,\theta,\phi)$, $h_{11}=h_{11}(t,r,\theta,\phi)$, $h_{10}=h_{10}(t,r,\theta,\phi)$, and $h_{22}=h_{22}(t,r,\theta,\phi)$ represent the components of the metric perturbation and depend on the coordinates $t$, $r$, $\theta$, and $\phi$. 

To investigate the anisotropy effects on the fluid pulsation mode emitted by compact stars, following \cite{thorne1967} and adopting the Regge and Wheeler \cite{regge1957} gauge, the metric perturbation $h_{\mu\nu}$ for a given even-parity spherical harmonic function $Y_{\ell m}(\theta,\phi)$ is given by 
%
%\begin{equation}
%h_{\mu\nu}=
%\begin{pmatrix}
%r^{\ell} {\tilde H}e^{i\omega t}e^{2\Psi} & i\omega r^{\ell+1}{\tilde H}_1e^{i\omega t} & 0 & 0\\
%i\omega r^{\ell+1}{\tilde H}_1e^{i\omega t} & r^{\ell}{\tilde H}e^{i\omega t}e^{2\Lambda} & 0 & 0\\
%0 & 0 & r^{\ell+2}{\tilde K}e^{i\omega t} & 0\\
%0 & 0 & 0 & r^{\ell+2}{\tilde K}e^{i\omega t}\sin^2\theta
%\end{pmatrix}
%Y_{\ell m}.
%\end{equation}
%Here, the functions ${\tilde H}$, ${\tilde H}_1$, and ${\tilde K}$ depend on $r$ and $\omega$ represents the eigenfrequency of oscillation. 
%
\begin{equation}\label{metric_perturbation}
h_{\mu\nu}=
\begin{pmatrix}
H e^{2\Psi} & H_1 & 0 & 0\\
H_1  & H e^{2\Lambda} & 0 & 0\\
0 & 0 & Kr^2 & 0\\
0 & 0 & 0 & Kr^2\sin^2\theta
\end{pmatrix}
Y_{\ell m},
\end{equation}
with $H$, $H_1$, and $K$ depending on $t$ and $r$. Through the non-zero components of the perturbed Einstein field equation \eqref{perturbed_FE} and the perturbed components of the conservation of the energy-momentum tensor ($\delta\left(\nabla_{\nu}T^{\nu}_1\right)=0$ and $\delta\left(\nabla_{\nu}T^{\nu}_2\right)=0$), the nonradial oscillation equations for a general anisotropic profile can be obtained; see Eqs.~\eqref{var_H_der}-\eqref{delta_T2nu}. Since we concentrate our attention on normal modes, following Ref.~\cite{detweiler1983,detweiler1985}, we employ $H=r^{\ell} {\tilde H}e^{i\omega t}$, $H_1=i\omega r^{\ell+1}{\tilde H}_1e^{i\omega t}$, $K=r^{\ell}{\tilde K}e^{i\omega t}$, $W=r^{\ell+1}{\tilde W}e^{i\omega t}$, and $V=r^{\ell}{\tilde V}e^{i\omega t}$; where the functions ${\tilde H}$, ${\tilde H}_1$, ${\tilde K}$, ${\tilde W}$, and ${\tilde V}$ depend on $r$, and $\omega$ represents the oscillation eigenfrequency. By using these five equalities and introducing the function ${\tilde X}$ defined by 
\begin{equation}\label{eq_X_tilde}
{\tilde X}=\left(1+\frac{\partial\sigma}{\partial p_r}\right)^{-1}\left[\omega^2(\rho+p_r+\sigma){\tilde V}e^{-\Psi}-\left(\rho+p_r\right)\frac{{\tilde H}}{2}e^{\Psi}+e^{\Psi+2\Lambda}{\tilde H}\frac{\partial\sigma}{\partial g_{11}}\right]-\frac{p_r'{\tilde W}}{re^{\Lambda-\Psi}},    
\end{equation}
equations \eqref{var_H_der}-\eqref{delta_T2nu} yield
%
%\begin{widetext}
\begin{eqnarray}
&&{\tilde H}_1'=\frac{e^{2\Lambda}}{r}\left[{\tilde H}+{\tilde K}+16\pi(\rho+p_r+\sigma){\tilde V}\right]-\frac{{\tilde H}_1}{r}\left[\ell+1+\frac{2me^{2\Lambda}}{r}+4\pi r^2(p_r-\rho)e^{2\Lambda}\right],\label{eq_H1_tilde_der}\\
&&{\tilde K}'=\frac{{\tilde H}}{r}+\frac{\ell(\ell+1)}{2r}{\tilde H}_1-\frac{{\tilde K}}{r}\left(\ell+1-\Psi'r\right)+8\pi\left(\rho+p_r\right)\frac{e^{\Lambda}{\tilde W}}{r},\label{eq_K_tilde_der}\\
&&{\tilde W}'=-(\ell+1)\frac{{\tilde W}}{r}+re^{\Lambda}\left[\frac{e^{-\Psi}{\tilde X}}{p_r+\rho}\frac{d\rho}{dp_r}-\ell(\ell+1)\frac{(p_r+\rho+\sigma)}{(p_r+\rho)}\frac{{\tilde V}}{r^2}-\frac{{\tilde H}}{2}\right.\nonumber\\
&&\left.-\frac{2\sigma {\tilde W}}{e^{\Lambda}r^2(p_r+\rho)}-\frac{(p_r+\rho+\sigma)}{(p_r+\rho)}{\tilde K}\right],\label{eq_W_tilde_der}\\
&&{\tilde X}'={\tilde W}\frac{e^{\Psi}}{r}\left[(p_r+\rho)\left[-4\pi(\rho+p_r){e^{\Lambda}}-\omega^2{e^{\Lambda-2\Psi}}+\left[\frac{e^{-\Lambda}\Psi'}{r^2}\right]'r^2\right]+\frac{2}{r}\left[\frac{3\sigma}{r}+p_r'\frac{\partial\sigma}{\partial p_r}-\sigma'\right.\right.\nonumber\\
&&\left.\left.-2\sigma\Psi'+\frac{2\sigma^2}{r(p_r+\rho)}\right]e^{-\Lambda}\right]-{\tilde H}_1\frac{e^{\Psi}}{2}\left[(p_r+\rho)\left[\frac{\ell(\ell+1)}{2r}+{e^{-2\Psi}r}\omega^2\right]+\frac{\ell(\ell+1)\sigma}{r}\right]\nonumber\\
&&+{\tilde H}\left[\frac{e^{\Psi}}{2}\left(p_r+\rho\right)\left[\Psi'-\frac{1}{r}\right]-\frac{2}{r}e^{\Psi+2\Lambda}\frac{\partial\sigma}{\partial g_{11}}\right]+{\tilde K}\left[(p_r+\rho)\frac{e^{\Psi}}{2}\left[\frac{1}{r}-3\Psi'\right]+\sigma e^{\Psi}\left[\frac{3}{r}\right.\right.\nonumber\\
&&\left.\left.-2\Psi'+\frac{2\sigma}{r(p_r+\rho)}\right]\right]-{\tilde X}\left[\frac{\ell}{r}+\frac{d\rho}{dp_r}\frac{2\sigma}{r(p_r+\rho)}-\frac{2}{r}\frac{\partial\sigma}{\partial p_r}\right]+{\tilde V}\frac{e^{\Psi}}{r^2}\ell(\ell+1)\left[-(p_r+\rho)\Psi'\right.\nonumber\\
&&\left.+\frac{2\sigma}{r}\right]\left[1+\frac{\sigma}{p_r+\rho}\right],\label{eq_X_tilde_der}\\
&&{\tilde H}\left[3m+\frac{1}{2}(\ell+2)(\ell-1)r+4\pi p_rr^3\right]={\tilde H}_1\left[\omega^2r^3e^{-2(\Psi+\Lambda)}-\frac{\ell(\ell+1)}{2}\left(m+4\pi r^3p_r\right)\right]\nonumber\\
&&+{\tilde K}\left[\frac{1}{2}(\ell+2)(\ell-1)r-\omega^2r^3e^{-2\Psi}-\frac{e^{2\Lambda}}{r}\left(m+4\pi r^3p_r\right)\left(3m-r+4\pi r^3p_r\right)\right]-8\pi e^{-\Psi}{\tilde X}r^3\nonumber\\
&&-16\pi e^{-\Lambda}\sigma {\tilde W}r.\label{eq_h}
\end{eqnarray}
%\end{widetext}
Here, the anisotropic profile employed depends on the fluid and spacetime variables of the form $\sigma=\sigma(p_r,g_{11})$ with $p_r=p_r(\rho)$; see, e.g., \cite{doneva2012,arbanil2023}. Eqs.~\eqref{eq_X_tilde}-\eqref{eq_h} can be reduced to the fourth-order system of linear equations presented in \cite{detweiler1985} by taking ${\tilde H}\rightarrow-{H}$, ${\tilde H}_1\rightarrow-{H}_1$, ${\tilde K}\rightarrow-{K}$, and $\sigma=0$; see also \cite{jun-li2011}. In addition, it is important to point out that Eqs.~\eqref{eq_W_tilde_der} and \eqref{eq_X_tilde_der} differ from the respective equations derived in Ref.~\cite{modal_2024, mondal2025} in all terms where the anisotropic factor appears; however, these equations are in concordance with those reported in Ref.~\cite{2024PhRvD.110h3020L,2026PhRvD.113d3025G,2026arXiv260106962Y}. The difference with \cite{modal_2024, mondal2025} occurs because, in the anisotropic case, the first law of thermodynamics needs to consider that the work done on the fluid varies with direction.

To solve Eqs.~\eqref{eq_X_tilde}-(\ref{eq_h}), some conditions at the center ($r=0$) and at the surface of the star ($r=R$) are required. At the center of the star, as in the isotropic case \cite{detweiler1983,detweiler1985}, regularity conditions of the perturbative variables are imposed. These conditions are found by expanding the fluid variables and the spacetime perturbations by Taylor power series near $r=0$. These expansions, together with the perturbation equations (\ref{eq_X_tilde})-(\ref{eq_h}) reduced through these expansions, are shown in Appendix \ref{Appendix_B}. In turn, outside the stellar structure configuration $r\geq R$, the fluid variables $p_r, \rho$, and $\sigma$ and the fluid perturbation quantities $W$ and $V$ are zero and $m=M$. Moreover, following \cite{sotani2001}, we replace the variables ${\tilde H}_1$ and ${\tilde K}$ with the new ones $Z$ and $dZ/dr^*$ through the equalities:
\begin{eqnarray}
&&\hspace{-1cm}{\tilde K}=\frac{n(n+1)r^2+3nMr+6M^2}{r^{\ell+2}(nr+3M)}Z+\frac{1}{r^{\ell+2}}\frac{dZ}{dr^*},\label{new_variables}\\
&&\hspace{-1cm}{\tilde H}_1=\frac{nr^2-3nMr-3M^2}{r^{\ell+1}(r-2M)(nr+3M)}Z+\frac{r^{1-\ell}}{(r-2M)}\frac{dZ}{dr^*},\label{new_variables2}
\end{eqnarray}
giving rise to the Zerilli equation \cite{zerilli1970,fackerell1971,chandrasekhar_detweiler1975}. In Eqs.~\eqref{new_variables} and \eqref{new_variables2}, $n=(\ell-1)(\ell+2)/2$ and $r^*$ is the so-called ``tortoise'' coordinate, which is related to the radial coordinate $r$ through the relation:
\begin{equation}
    r^*=r+2M\ln\left(\frac{r}{2M}-1\right).
\end{equation}

At the star's surface ($r=R$), the Lagrangian perturbation $\Delta p_r$ must vanish, i.e.,
\begin{equation}
\Delta p_r=-r^{\ell}e^{-\Psi}{\tilde X}=0,
\end{equation}
thus implying ${\tilde X}(r=R)=0$. {Each mode solution is uniquely associated with a specific ($\ell, \omega$) pair; however, not every combination of ($\ell, \omega$) corresponds to a valid mode solution, since the spectrum is discrete.}
In order to consider quadrupolar modes (linked to gravitational radiation), we set $\ell = 2$.
%For a given set of $\ell$ and $\omega$, all the boundary conditions are satisfied by only one solution. 

%\subsection{Tidal deformability}

\subsection{Tidal deformability}\label{subsection2}

{The investigation of tidal Love numbers has emerged as a key aspect in understanding binary systems consisting of compact stars \cite{Flanagan2008,Chatziioannou2020}. In these systems, the gravitational pull exerted by one star induces tidal distortions in its companion. Such deformations, arising from external tidal forces, are characterized by the dimensionless tidal deformability parameter $\Lambda$, which is defined as \cite{Hinderer2008,Flanagan2008}
\begin{equation}\label{tidal_deformability}
    \Lambda = \frac{2k_2}{3C^5}.
\end{equation}
In this equation, $C = M/R$ represents the stellar compactness, and $k_2$ is the quadrupolar Love number, which can be written as \cite{Hinderer2008}
\begin{align}\label{k2}
{k_2} =&\ \frac{8C^5}{5}(1-2C)^2[2+2C(y_R-1)-y_R]\times\{2C [6-3y_R+3C(5y_R-8)]+4C^3[13-11y_R \nonumber \\
&+C(3y_R-2)+2C^2(1+y_R)]+3(1-2C)^2\times [2-y_R+2C(y_R-1)]\ln(1-2C)\}^{-1},
\end{align}
where $y_R = y(r=R)$. The function $y(r)$ follows the Riccati-type differential equation
\begin{equation}\label{riccati_eq}
y'r + y^2 + y\left(K_0 r - 1\right) + K_1 r^2 = 0,
\end{equation}
with coefficients
\begin{align}
    K_0 =&\ \frac{2m}{r^2}e^{2\Lambda} + 4\pi e^{2\Lambda}(p_r-\rho)r + \frac{2}{r}, \label{C_0_y}  \\
    K_1 =&\ 4\pi e^{2\Lambda}\left[4\rho + 8p_r + 4\sigma + \frac{p_r+\rho}{Ac_s^2}(c_s^2+1)\right]- \frac{6}{r^2}e^{2\Lambda} - 4\Psi'^2, \label{C_1_y}
\end{align}
where $c_s^2 = \frac{dp_r}{d\rho}$ and $A = \frac{dp_t}{dp_r}$, where $p_t$ denotes the tangential pressure, defined as $p_t=p_r+\sigma$. Additional information can be found in Ref.~\cite{arbanil2023} and the references therein.

%For strange quark stars, where the surface energy density remains nonzero, it becomes necessary to introduce a correction to $y(R)$. Because of the discontinuity present in the energy distribution, the boundary condition is adjusted as follows: \cite{wang2019, li2020, zhou2018, lourenco2021}
%
%\begin{equation}
%y_R \longrightarrow y_R - \frac{4\pi R^3 \rho_s}{M},
%\end{equation}
%with $\rho_s$ being the difference in energy density across the stellar surface.
}

%%%%%%%%%%%%%%%%%%%%%%%%%%%%%%%%%%%%%%%%%%%%%%%%%%%%%%%%%%%%%
\section{Equation of state and anisotropic profile}\label{section3}
%%%%%%%%%%%%%%%%%%%%%%%%%%%%%%%%%%%%%%%%%%%%%%%%%%%%%%%%%%%%%

\subsection{Equation of state}

In the outer layers of a compact object, the composition of matter evolves from nuclei embedded in an electron sea in the outer crust to progressively more neutron-rich matter in its innermost regions. In these low density regions, the EOS has been derived from many-body calculations using phenomenological potentials, typically including two and three nucleon interactions \cite{PhysRevC.58.1804}. Recently, chiral effective field theory (CEFT) has provided a systematic framework for expanding nuclear forces at low momenta, naturally explaining the hierarchy among two-body, three-body, and weaker higher-body interactions \citep{Coraggio:2012ca,PhysRevC.85.032801,PhysRevC.87.014338,PhysRevC.82.014314,PhysRevC.86.054317,Tews:2012fj}. At present, microscopic calculations based on CEFT interactions allow a reliable determination of  the properties of NS matter up to the nuclear saturation density $n_0$.

In the inner regions of a NS, at moderately high densities, the EOS of cold quark matter can be obtained in the framework of pQCD. 
The resulting EOS depends on an unphysical parameter, namely the renormalization scale associated with the chosen scheme. Although this dependence decreases order by order in perturbation theory, it provides a practical estimate of the contributions from yet uncalculated higher-order terms and thus serves as a quantitative measure of the intrinsic theoretical uncertainty \cite{Kurkela_2014}.  To describe the EOS in this region, we use the numerical EOS derived in \cite{PhysRevD.81.105021} which is well represented by a fitting function for the pressure in terms of the baryon chemical potential $\mu_B$ \cite{Fraga_2014}. Fixing the strong coupling constant and the strange quark mass at arbitrary reference scales (using lattice and experimental data), the quark-matter EOS in $\beta$-equilibrium reads 
\begin{equation}
\label{EOS1}                                                                                                                                   
P_{QCD}(\mu_{B})=P_{SB}(\mu_{B}) \left[    c_{1} - \dfrac{a(X)}{(\mu_{B}/GeV)-b(X)}   \right],
\end{equation}
with $P_{SB} = \tfrac{3}{4\pi^2}(\mu_{B}/3)^{4}$, $a(X) = d_{1}X^{-\nu_{1}}$ and $b(X) = d_{2}X^{-\nu_{2}}$, 
where $X$ is a parameter proportional to the renormalization scale of the theory (typically varied from $1$ to $4$). The parameters are  $c_{1} = 0.9008$,   $d_{1}= 0.5034$,   $d_{2}= 1.452$,    $\nu_{1}= 0.3553$ and $\nu_{2}= 0.9101$.
These formulae reproduce the full three-loop pressure correctly, the quark number density and the speed of sound to per cent accuracy for baryon chemical potentials up to $6$\,[GeV] \cite{Kurkela_2014,Fraga_2014}.

To construct the EOS over the entire density range, we proceed as follows. For densities below $1.1n_0$,  the CEFT EOS of \cite{Tews:2012fj} is used. 
At baryon chemical potentials above $2.6$\,[GeV], where the relative uncertainty of the quark matter EOS is as large as the nuclear matter one at $n=1.1 n_0$, the expression of Eq.~\eqref{EOS1} is used. Between these two regions, it is assumed that the EOS is well approximated by an interpolating polytrope built from two monotropes of the form 
\begin{equation}
P_{i}(\mu_{B}) = k_{i}\left( n_{i}^{\gamma_{i} - 1} + \dfrac{\gamma_{i}-1}{k_{i}\gamma_{i}}(\mu_{B} - \mu_{B,i}) \right)^{\tfrac{\gamma_{i}}{\gamma_{i}-1}}, 
\end{equation}
where $k_{i}$ and $\gamma_{i}$ are the polytropic coefficient and the polytropic exponent, respectively. In this work  $i = 1,2$, which means the matching of two monotropes. Furthermore, $\mu_{B,i}$ and $n_{i}$ represent the baryon chemical potential and the baryon density, respectively. 
These functions are connected smoothly  matched together at a set of intermediate chemical potentials. We will use the representative EOS III given in \cite{Kurkela_2014} as the interpolating polytrope that approximates the EOS between the two aforementioned regions (see Table \ref{table4}). We note that EOS III satisfies the requirement that the EOS is able to support a $2\, M_\odot$ NS (to be in agreement with recent mass determinations using pulsar timing data \cite{demorest2010,antoniadis2013}).

\begin{table}
$$
\begin{array}{c||cccc}
n/n_0	&	p &	\rho	&	\mu	&c_s^2\\
\hline
\hline
 1.1 & 3.542 & 168.5 & 0.9775 & 0.15\\
 1.3 & 12.13 & 200.4 & 1.022  & 0.42\\
 1.5 & 34.81 & 234.5 & 1.122  & 0.95\\
 \hline
 1.7 & 42.24 & 270.9 & 1.151 & 0.19\\
 1.9 & 49.44 & 308.1 & 1.176 & 0.20\\
 2.1 & 56.96 & 346.2 & 1.200 & 0.20\\
 2.3 & 64.79 & 384.9 & 1.222 & 0.20\\
 2.5 & 72.90 & 424.4 & 1.243 & 0.21\\
 2.7 & 81.29 & 464.5 & 1.263 & 0.21\\
 2.9 & 89.94 & 505.2 & 1.283 & 0.21\\
 3.1 & 98.84 & 546.5 & 1.301 & 0.22\\
 3.3 & 108.0 & 588.5 & 1.319 & 0.22\\
 3.5 & 117.4 & 631.0 & 1.336 & 0.22\\
 3.7 & 127.0 & 674.0 & 1.353 & 0.22\\
 3.9 & 136.8 & 717.5 & 1.369 & 0.23 
\end{array}
$$
\caption{The representative EOS III. The baryon number density $n$ is given in units of the saturation density $n_0=0.16\,[\rm{fm}^{-3}]$, while the pressure $p$ and the energy density $\rho$ are given in $[\rm{ MeV/fm}^3]$. 
%$R$ (in $[{\rm km}]$) and $M$(in solar masses) stands for the radius and mass of a star with central density $n$, while 
The solid horizontal line indicates the transition between the two monotropes \cite{Kurkela_2014}.}
\label{table4}
\end{table}

\subsection{Anisotropic profile}

To describe the anisotropic profile, we employ the quasilocal anisotropic profile. As described in \cite{horvat2011}, it depends on the fluid and spacetime variables of the form $\sigma=\sigma(p_r,g_{11})$. Thus, this equation is given by:
\begin{equation}\label{anisotropic_eos}
\sigma=\alpha p_r\left(1-\frac{1}{g_{11}}\right)^a,
\end{equation}
where $\alpha$ is a dimensionless anisotropic constant, $a$ being a constant dimensionless parameter, and $g_{11}=e^{2\Lambda}$. The anisotropy model \eqref{anisotropic_eos} was used, e.g., with $a=1$, to analyze the effects of the anisotropy on the radial pulsation modes of polytropic stars \cite{horvat2011,arbanil_panotopoulos2022} and strange quark stars \cite{arbanil_malheiro_2016}, nonradial oscillation modes within the Cowling approximation of NSs \cite{doneva2012} and strange quark stars \cite{arbanil2023}, magnetic field structure \cite{folomeev2015}, and slowly rotating NSs \cite{silva2015,Pretel2022} and, with $a=2$, to investigate its influence on quasinormal modes \cite{2024PhRvD.110h3020L,2026arXiv260106962Y} and on the fundamental oscillation modes of NSs \cite{2026PhRvD.113d3025G}. In the present work, in order to compare our results with those reported in the literature (see, e.g., Refs.~\cite{doneva2012,arbanil2023}), we consider the range $-1 \leq \alpha \leq 1$. Furthermore, following Ref.~\cite{2026PhRvD.113d3025G}, we adopt $a = 2$. The values $\alpha = \pm 1$ are chosen as representative cases within a phenomenological framework. Since $\alpha$ is a dimensionless anisotropic constant controlling the magnitude of the anisotropy relative to the radial pressure, values of order unity ensure that the anisotropic contribution remains comparable to the local pressure. Moreover, the choice $\alpha = \pm 1$ allows us to explore both regimes of anisotropy, namely $\sigma > 0$ and $\sigma < 0$, in a symmetric manner around the isotropic case $\alpha = 0$. This behavior can be physically understood from the anisotropic contribution to the hydrostatic equilibrium: positive anisotropy $(p_t>p_r)$ gives rise to an effective outward force that partially counteracts gravity, whereas negative anisotropy $(p_t<p_r)$ produces an effective inward force, enhancing the gravitational attraction and reducing the stellar support against collapse \cite{Hernandez2004, PretelSBD2022}.

%In the present work, to compare our results with some results reported in the literature -following, e.g., Ref. \cite{doneva2012,arbanil2023}- we consider $-1\leq\alpha\leq1$ and -based on, e.g., Ref.\cite{2026PhRvD.113d3025G}- we employ $a=2$.

For the anisotropic profile \eqref{anisotropic_eos}, we note that 
\begin{eqnarray}
&&\frac{\partial\sigma}{\partial g_{11}}=\frac{2\alpha p_r}{g_{11}^2}\left(1-\frac{1}{g_{11}}\right),\quad \frac{\partial\sigma}{\partial p_r}=\alpha\left(1-\frac{1}{g_{11}}\right)^2,\quad{\rm and}\quad\frac{\partial\sigma}{\partial\rho}=\frac{\partial p_r}{\partial \rho}\frac{\partial\sigma}{\partial p_r}.
\end{eqnarray}
It is important to highlight these factors since they appear in Eqs.~\eqref{eq_X_tilde}--\eqref{eq_h}. In this way, the set of equations used to analyze nonradial oscillations is complete.

%%%%%%%%%%%%%%%%%%%%%%%%%%%%%%%%%%%%%%%%%%%%%%%%%%
\section{Numerical Methods}\label{sectio_numerical_method}
%%%%%%%%%%%%%%%%%%%%%%%%%%%%%%%%%%%%%%%%%%%%%%%%%%

In this section, we briefly summarize the methods used to compute the nonradial oscillation modes and the tidal deformability of NSs.

\subsection{Nonradial oscillation modes} 

To determine the fundamental mode frequency, we choose a suitable trial Newtonian value  $\omega_{Newt} = \sqrt{(M/R^{3})2l(l-1)/(2l+1)}$ as a starting point, as done in Ref.~\cite{detweiler1983}. Then we start the integration process within the stellar interior. First, we construct three linearly independent solutions that satisfy the regularity conditions at the star's center and numerically integrate them outward from $r = 0$ to the midpoint at $R/2$. Next, we construct two additional linearly independent solutions that meet the boundary conditions at the stellar surface, and integrate them inward from $r = R$ to $R/2$. Finally, at $R/2$, we combine these five solutions, in a way that ensures the resulting function satisfies the boundary conditions at both the center and the surface, completing the computation of the mode within the star. 

The next step is to solve the Zerilli equation in the exterior region of the star. To accomplish this, we first determine the boundary values at the stellar surface for both the Zerilli function and its radial derivative. These can be extracted from the values of the metric perturbation functions $H(R)$ and $K(R)$ obtained through the interior integration. With these boundary conditions at $r = R$, the Zerilli equation is integrated outward in terms of the tortoise coordinate $r^{*}$. In the asymptotic regime, where $r^{*} \to \infty$, the general solution of the Zerilli equation can be represented as a linear combination of ingoing and outgoing waves, as can be found in Ref.~\cite{jun-li2011}.

To determine the quasi-normal mode frequencies, we have to integrate the Zerilli equation from the surface of the star to a sufficiently large radial coordinate, typically taken as $r_{\infty} \approx 50\, \omega^{-1}$ (during that integration, $\omega$ is treated as a real parameter in both the interior and exterior regions). The physical boundary condition at infinity imposes that only outgoing gravitational waves should be present. Therefore, the problem of finding quasi-normal mode frequencies reduces to locating the complex roots as done in~\cite{vasquez2014}.

This procedure is repeated iteratively: the real part of the estimated root is used as the new trial input in the next integration cycle. The iterations continue until the real part of the frequency converges to within one part in $10^8$ between successive steps. Additionally, once convergence is reached, the imaginary part of the frequency $\mathrm{Im}(\omega)$ yields the damping timescale of the mode.

%{\it For tidal deformability:} {\bf The deformability parameter is obtained by solving the Riccati equation Eq.~(\ref{riccati_eq}), coupled with the TOV equations Eqs.~(\ref{tov},\ref{conservatio_mass},\ref{phi_eq}), from the center of the star ($r = 0$) to its surface ($r = R$), where the pressure is required to vanish. Once the system is solved, the value of $y(r = R)$ (at the surface) is determined, which, when substituted into equation \eqref{k2}, yields the dimensionless deformability parameter $\Lambda$ via \eqref{tidal_deformability}.}

\subsection{Tidal deformability}

{To examine how anisotropy influences the tidal deformability of NSs, we numerically integrate the stellar structure equations \eqref{tov}-\eqref{phi_eq} together with the tidal deformability equation \eqref{riccati_eq} from the center of the star ($r = 0$) to its surface ($r = R$). Specifically, Eqs.~\eqref{tov}-\eqref{phi_eq} are first solved using the fourth-order Runge--Kutta method for various choices of the anisotropy parameter $\alpha$ and central energy density $\rho_c$. This step provides the radial profiles for $p_r$, $\sigma$, $\rho$, $m$, and $\Psi$. Next, Eq.~\eqref{phi_eq} is handled via the shooting method: an initial trial for $\Psi_c$ at the center is supplied, and if the resulting solution does not satisfy the boundary condition given in equation \eqref{phi_eq}, $\Psi_c$ is iteratively adjusted until convergence is achieved. After the consistent $\Psi_c$ is determined, the tidal deformability equation \eqref{riccati_eq} is then integrated outward from $r = 0$ to $r = R$, ensuring a self-consistent solution for each selected pair of $\alpha$ and $\rho_c$. Once the deformability function $y(r)$ is known, the value of $y(r = R)$ is determined by means of Eq.~\eqref{k2}, which finally allows us to calculate the dimensionless deformability parameter \eqref{tidal_deformability}.}

%%%%%%%%%%%%%%%%%%%%%%%%%%%%%%%%%%%%%%%%%%%%%%%%%%%%
\section{Anisotropic effects on the equilibrium, oscillation frequency, and tidal deformability}\label{results}
%%%%%%%%%%%%%%%%%%%%%%%%%%%%%%%%%%%%%%%%%%%%%%%%%%%%

\subsection{Equilibrium configurations and oscillation frequencies}

\begin{figure}[h]
    \centering
    \includegraphics[width=0.49\textwidth]{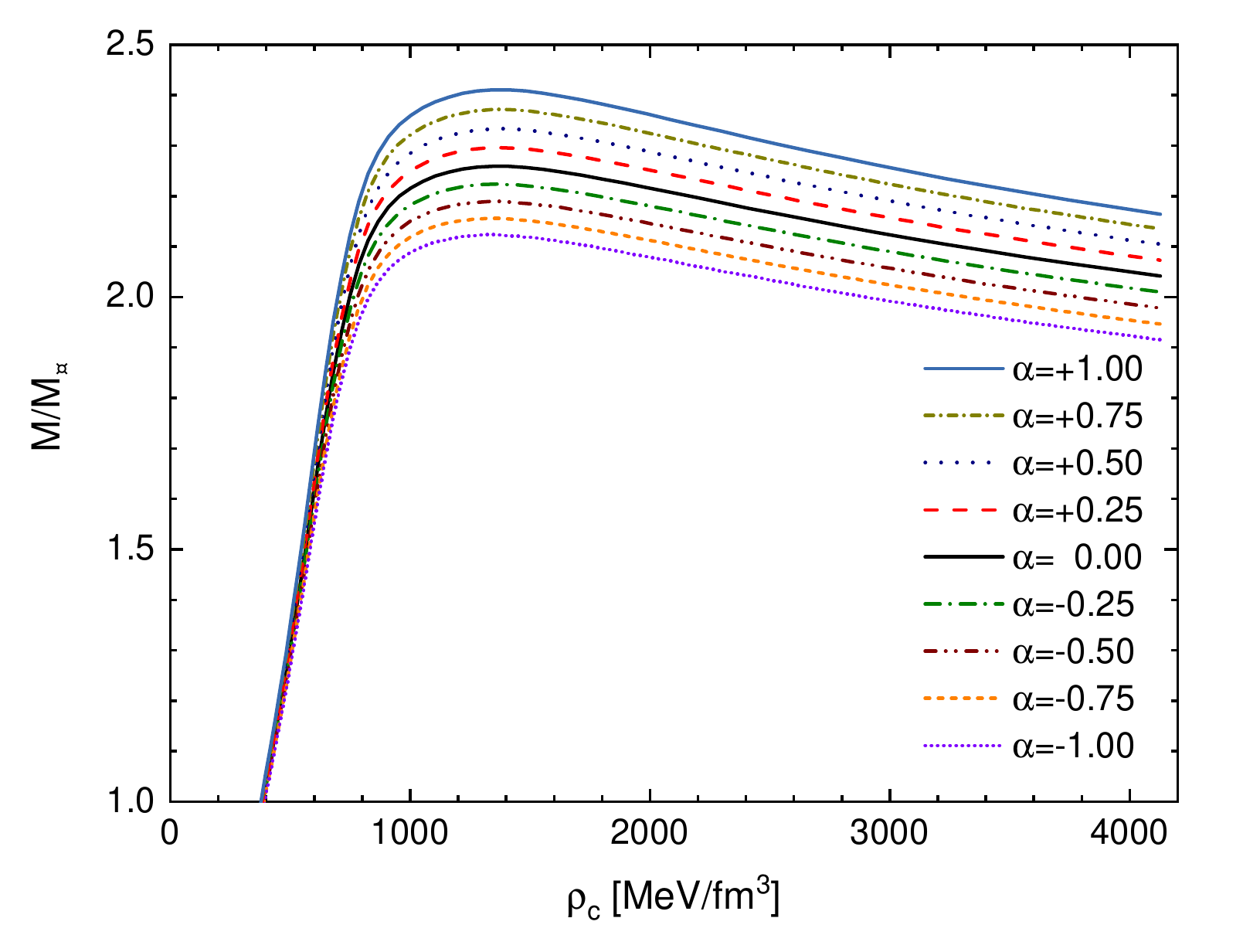}
    \includegraphics[width=0.49\textwidth]{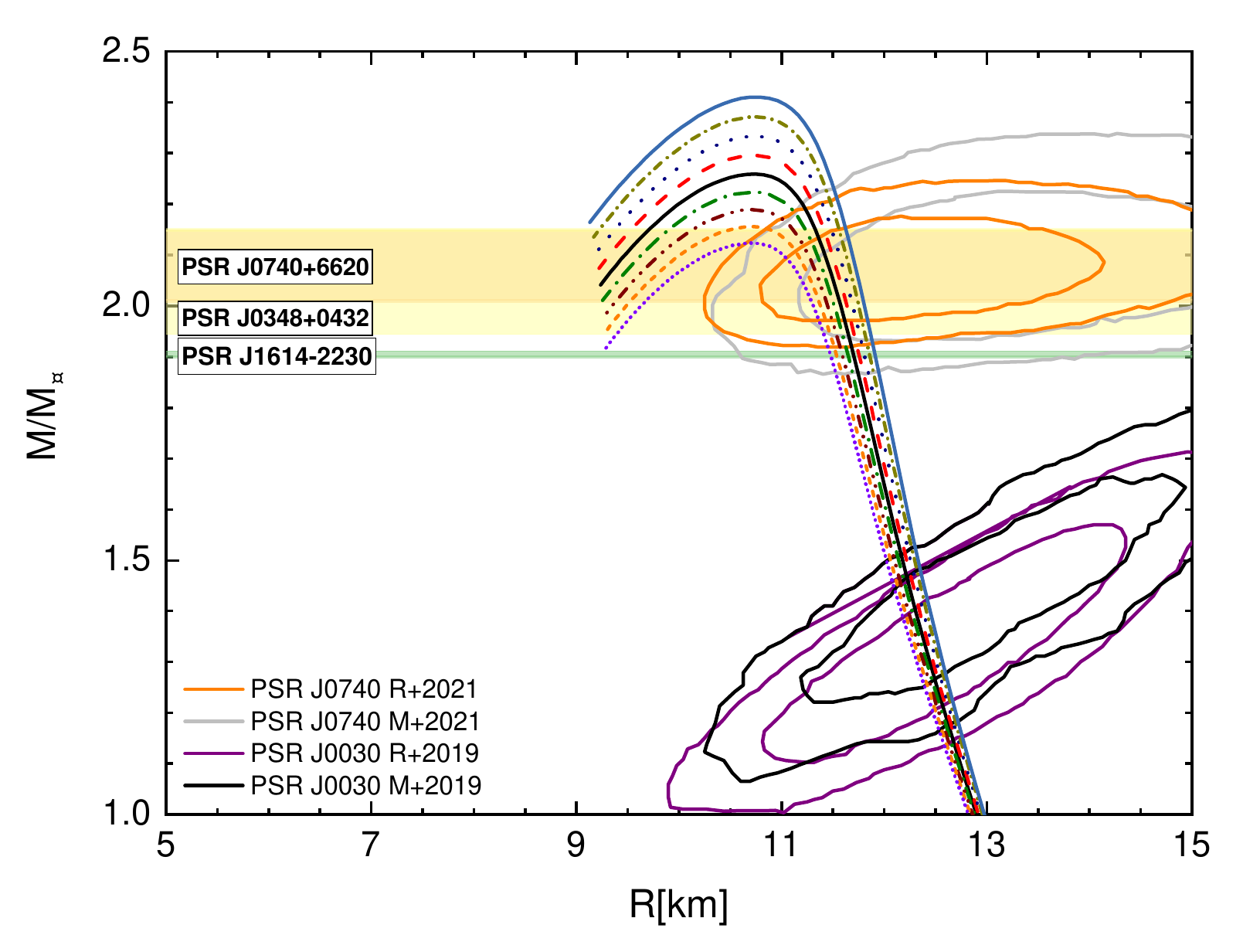}
    \caption{The total gravitational mass, in solar masses, as a function of the central energy density and of the stellar radius is shown in the left and right panels, respectively. In both panels, nine different values of the anisotropic parameter $\alpha$ are considered. In the right panel, the purple and black contours correspond to the mass–radius constraints of PSR J$0030+0451$ \cite{riley2019,miller2019}, while the orange and gray contours represent those of PSR J$0740+6620$ \cite{riley2021,miller2021}, as reported by the NICER collaboration. For each source, the inner and outer contours denote the $68\%$ and $95\%$ confidence intervals, respectively. The outer contours, therefore, encompass a higher probability of containing the true values but with lower precision, whereas the inner contours provide tighter constraints with lower statistical coverage. In addition, the horizontal bands in the right panel represent the observational mass constraints from well-measured massive pulsars, namely PSR J$0740+6620$, PSR J$0348+0432$, and PSR J$1614-2230$. The mass of PSR J$0740+6620$ was initially reported in \cite{cromartie2020} and later refined in \cite{fonseca2021}, while that of PSR J$1614-2230$ was first measured in \cite{demorest2010} and subsequently updated in \cite{fonseca2016}. These bands indicate the allowed mass ranges and provide an additional consistency check for the theoretical mass–radius relations.} 
    \label{fig:Mrho_MR}
\end{figure}

In Figure \ref{fig:Mrho_MR} we present the gravitational mass in units of solar masses as a function of the central energy density (left panel) and the radius (right panel) for nine different anisotropic parameters. We consider central energy densities in the range $400$–$4200\,[\rm MeV/fm^3]$. As shown in the left panel, overall, the stellar mass follows the expected trend of spherical NSs with increasing central energy density, i.e., it increases monotonically up to a maximum-mass configuration. Beyond this point, the mass decreases with further increases in $\rho_c$. Similarly, in the right panel, all the mass-radius curves follow the expected trend for NSs, i.e., as the central density increases, the mass increases while the radius simultaneously decreases, up to the maximum-mass configuration. At this point, the curves bend counterclockwise, and the mass subsequently decreases with further changes in the radius. In this panel, the mass-radius relation reported by NICER from compact stars PSR J$0030+0451$ \cite{riley2019,miller2019} and PSR J$0740+6620$ \cite{riley2021,miller2021} are also presented. The corresponding bands of the $2M_{\odot}$ pulsars PSR J$0740+6620$ \cite{cromartie2020,fonseca2021}, PSR J$0348+0432$ \cite{antoniadis2013}, and PSR J$1614-2230$ \cite{demorest2010,fonseca2016}, respectively marked by light yellow, light orange, and light green colors, are also shown in the mass-radius diagram. 

%{In such figure, since the central pressure becomes extremely high or low for large and small values of the anisotropic parameter, respectively, it is impossible to carry out numerical calculations with high central energy densities. Thus, we consider all equilibrium configurations where excellent numerical precision is found. Therefore, in such situations, the approach is to get as close as possible to the value corresponding to the maximum mass.  In all mass against central energy density curves, the mass increases monotonically with $\rho_c$ until values near the maximum mass values.} 

\begin{figure}[h]
    \centering
    \includegraphics[width=0.49\textwidth]{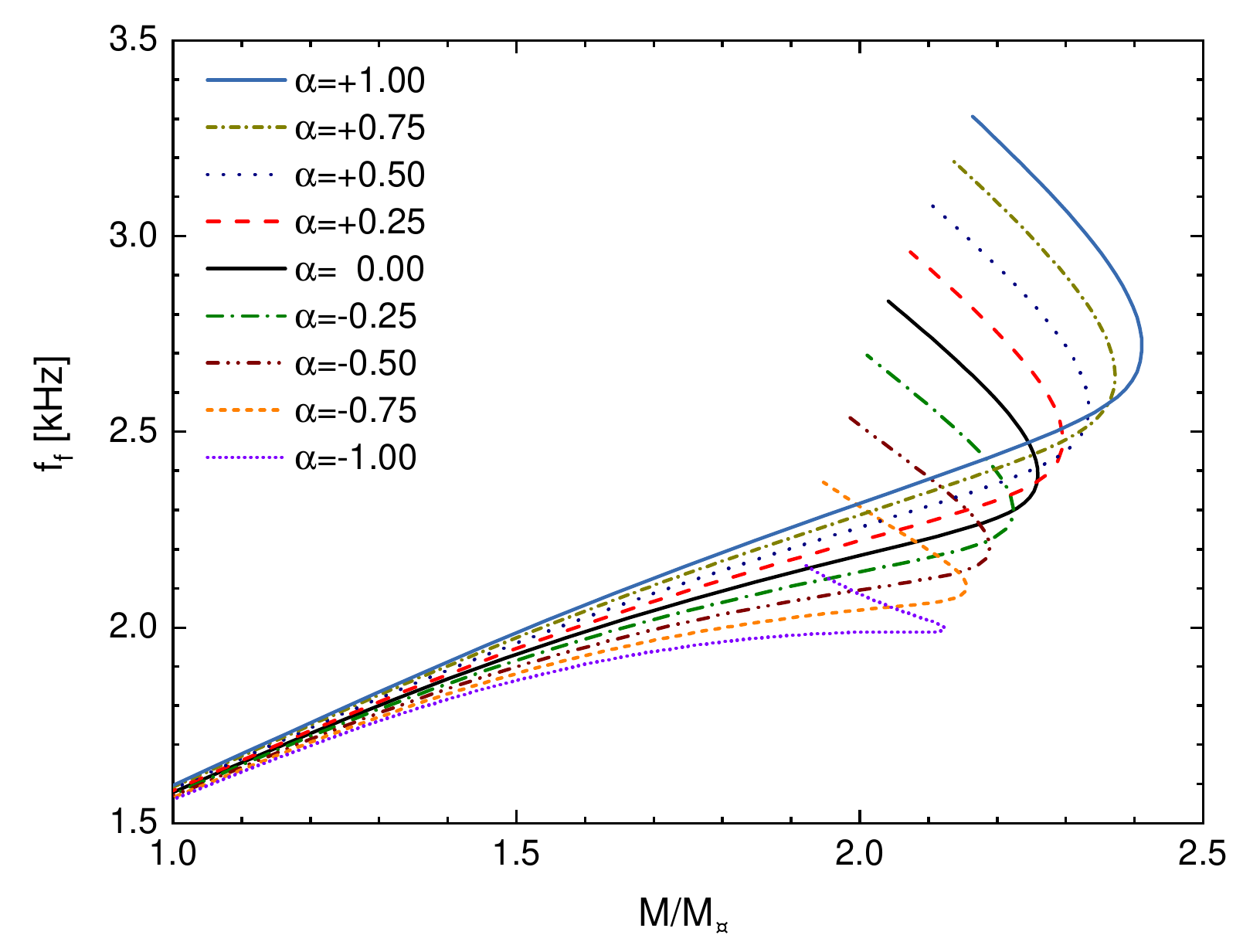}
    \includegraphics[width=0.49\textwidth]{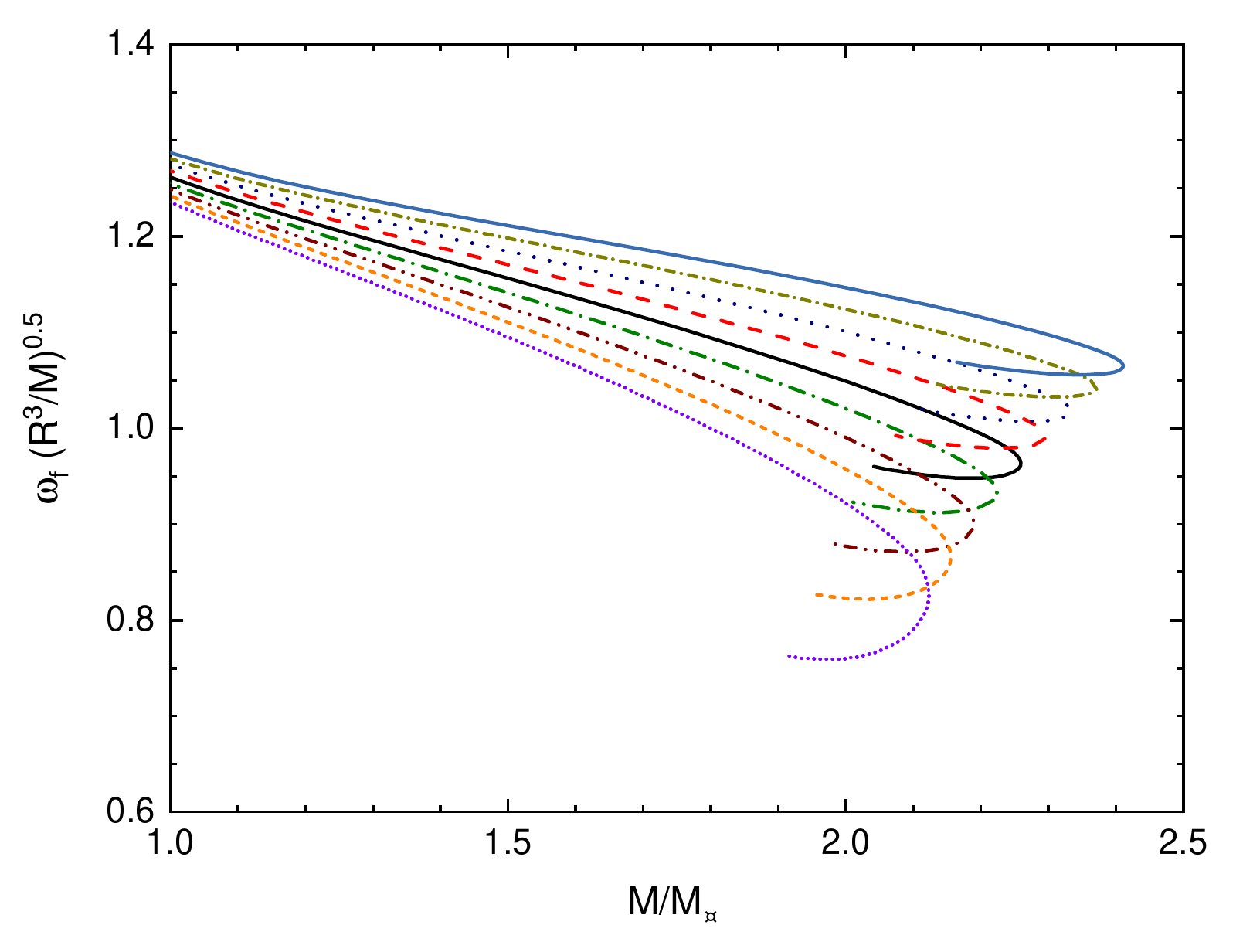}
    \caption{The $f$-mode frequency of oscillations and the normalized $\omega_f$ with the average density $(M/R^3)^{0.5}$ against the total mass are presented on the left and right panels, respectively. In both panels, nine different anisotropic parameters $\alpha$ are used.}\label{fig:fmodes_OmegaM}
\end{figure}

The effect of anisotropic pressure on mass and radius can also be observed in Fig.~\ref{fig:Mrho_MR}. Upon examining the maximum mass and its corresponding total radius, it becomes evident that the mass varies significantly with changes in $\alpha$; however, the total radius exhibits only a small variation. As $\alpha$ increases (decreases), the stellar configurations attain higher (lower) masses to achieve equilibrium. Moreover, the increase (decrease) in mass with $\sigma$ can be understood by noting the change of $\sigma$ with the parameter $\alpha$. Greater (Lesser) anisotropy $\sigma$ supports higher (lower) fluid pressure, enabling stars with larger (smaller) masses to resist collapse. 

In addition, from the mass–radius curves that intersect the NICER confidence regions and satisfy the heavy-pulsar mass constraints, it can be inferred that the model under consideration is consistent with current observational data. The agreement between the theoretical configurations and these observational bounds indicates that the model reproduces both the inferred masses and radii within the corresponding uncertainties. This comparison allows us to constrain the permissible range of the model’s physical parameters, thereby providing additional restrictions on the degree of anisotropy and, more generally, on the equation of state of the stellar fluid.

The $f$-mode nonradial oscillation frequency and the eigenfrequency normalized by the square root of the average density, $(\sqrt{M/R^3})$, as functions of the total mass, are shown in the left and right panels of Fig.~\ref{fig:fmodes_OmegaM}, respectively, for selected values of the anisotropy parameter $\alpha$.

To clarify the impact of the differences in the perturbation equations with respect to Refs.~\cite{modal_2024, mondal2025}, we perform a qualitative comparison between our results and those reported in these works. We find that the deviations in the $f$-mode frequencies are typically of the order of $\sim 3\text{--}5\%$ for canonical NS masses ($M \sim 1.4\,M_\odot$), increasing up to $\sim 10\text{--}15\%$ in the high-mass regime, close to the maximum-mass configuration. More specifically, Refs.~\cite{modal_2024, mondal2025} exhibit a more pronounced non-linear behavior of the $f$-mode frequency as a function of mass, including a crossing of the curves corresponding to different values of the anisotropy parameter. In contrast, our results display a smoother and more monotonic dependence, with no significant crossing behavior within the explored parameter space. These differences indicate that the treatment of anisotropy in the perturbation equations leads to systematic quantitative deviations, particularly for more massive configurations where relativistic effects are more relevant. Although the overall trends remain consistent, the observed discrepancies suggest that the detailed implementation of anisotropy in the perturbative framework can have a non-negligible impact on the predicted oscillation spectrum.

In the left panel of Fig. \ref{fig:fmodes_OmegaM}, all curves show that the $f$-mode frequency increases monotonically with the total mass up to the maximum-mass configuration. After this point, the curve turns counterclockwise, so that the fundamental frequency of the oscillation begins to increase with the decrease in total mass. The influence of anisotropy on the fluid pulsation mode is also evident. From the $f$-mode frequency versus $M/M_{\odot}$ curves, in certain mass ranges, the $f$-mode frequency increases or decreases in correspondence with an increase or decrease in $\alpha$.

In the right panel of Fig. \ref{fig:fmodes_OmegaM}, in all the curves presented, up to the point of maximum mass, the relationship between $\omega_f\sqrt{R^3/M}$ and $M/M_{\odot}$ produces an almost linear behavior. Beyond the maximum mass value, the curve bends clockwise and $\omega_f\sqrt{R^3/M}$ starts to increase with $M/M_{\odot}$. As noted in \cite{arbanil_malheiro_2016}, variations in the oscillation frequency are linked to changes in the radial fluid pressure induced by anisotropy.

%%%%%%%%%%%%%%%%%%%%%%%%%%%%%%%%%

\begin{table}[tbp]
\centering
%\begin{ruledtabular}
\begin{tabular}{ccccccc}
\hline
$\rho_c\,[\rm MeV/fm^3]$ & $\alpha$ & $M/M_{\odot}$ & $R\,[\rm km]$ & $M/R$ & $f_f\,[\rm kHz]$& $\omega_f\,\sqrt{R^3/M}$\\\hline
      & $-1.0$ & $1.0171$ & $12.792$ & $0.1174$ & $1.5726$ & $1.2305$\\
      & $-0.5$ & $1.0277$ & $12.815$ & $0.1184$ & $1.5902$ & $1.2413$\\
$400$ & $0.0$  & $1.0386$ & $12.838$ & $0.1194$ & $1.6082$ & $1.2523$\\
      & $+0.5$ & $1.0499$ & $12.863$ & $0.1205$ & $1.6268$ & $1.2634$\\
      & $+1.0$ & $1.0612$ & $12.888$ & $0.1216$ & $1.6458$ & $1.2749$ \\ \hline
      & $-1.0$ & $1.5498$ & $11.954$ & $0.1914$ & $1.8863$ & $1.0802$\\
      & $-0.5$ & $1.5819$ & $11.995$ & $0.1947$ & $1.9414$ & $1.1061$\\
$600$ & $0.0$  & $1.6159$ & $12.038$ & $0.1982$ & $1.9989$ & $1.1328$\\
      & $+0.5$ & $1.6519$ & $12.083$ & $0.2019$ & $2.0586$ & $1.1602$\\ 
      & $+1.0$ & $1.6902$ & $12.130$ & $0.2057$ & $2.1187$ & $1.1876$\\
\hline
\end{tabular}
\caption{\label{table}{The mass $M/M_{\odot}$, radius $R$, compactness $M/R$, $f$-mode oscillation frequency, and eigenfrequency normalized with the mean density $\omega_f\,\sqrt{R^3/M}$ for NSs with two different central energy densities and different values of $\alpha$. } } 
%\end{ruledtabular}
\end{table}

\begin{table}[tbp]
\centering
%\begin{ruledtabular}
\begin{tabular}{ccccccc}
\hline
$\alpha$ & $M_{\rm max}/M_{\odot}$ & $\rho_c\,[\rm MeV/fm^3]$ & $R\,[\rm km]$ & $M/R$ & $f_f\,[\rm kHz]$& $\omega_f\,\sqrt{R^3/M}$\\\hline
$-1.0$ & $2.1223$ & $1328.3$ & $10.682$ & $0.2934$ & $1.9983$ & $0.8260$\\
$-0.5$ & $2.1884$ & $1345.3$ & $10.693$ & $0.3022$ & $2.2041$ & $0.8986$\\
$0.0$ & $2.2585$ & $1358.1$ & $10.707$ & $0.3115$ & $2.3867$ & $0.9597$\\
$+0.5$ & $2.3324$ & $1366.6$ & $10.728$ & $0.3210$ & $2.5534$ & $1.0133$\\
$+1.0$ & $2.4100$ & $1366.6$ & $10.740$ & $0.3313$ & $2.7082$ & $1.0590$\\ 
\hline
\end{tabular}
\caption{\label{table3}{Maximum mass $M/M_{\odot}$, along with corresponding central energy density $\rho_c$, radius $R$, compactness $M/R$, $f$-mode oscillation frequency, and eigenfrequency normalized with the mean density $\omega_f\,\sqrt{R^3/M}$ for five values of $\alpha$.
} } 
%\end{ruledtabular}
\end{table}

%{\bf 
Table \ref{table} presents the mass, radius, compactness, $f$-mode pulsation frequency, and the oscillation eigenfrequency normalized with the mean density, for compact objects for the central energy densities $400\,[\rm MeV/fm^3]$ and $600\,[\rm MeV/fm^3]$ and five values of anisotropic parameters $\alpha$. It can be noted that, when varying \(\alpha\) from \(-1.0\) to \(+1.0\), the mass, radius, compactness, \(f\)-mode frequency, and normalized oscillation eigenfrequency change, respectively, by approximately \(4.34\%\), \(0.75\%\), \(3.58\%\), \(4.65\%\), and \(3.61\%\) for \(\rho_c = 400\,[\rm MeV/fm^3]\). For \(\rho_c = 600\,[\rm MeV/fm^3]\), the corresponding relative variations are approximately \(9.06\%\), \(1.47\%\), \(7.47\%\), \(12.32\%\), and \(9.94\%\). These values were computed using the standard relative-difference formula with the \(\alpha=-1\) configuration as the reference. Thus, for a central energy density interval, we note that when $\alpha$ increases, the mass, radius, compactness, $f$-mode frequency, and normalized eigenfrequency $\omega_f$ rise. From these results, it is possible to identify which physical quantities are most affected by the anisotropy parameter \(\alpha\). In both density regimes considered, the stellar mass exhibits the largest relative variations as \(\alpha\) varies, whereas the radius is affected to a much lesser extent. Consequently, derived quantities such as the compactness and the oscillation frequencies, which depend on combinations of \(M\) and \(R\), naturally inherit this behavior.

In particular, the \(f\)-mode frequency and the normalized oscillation eigenfrequency show significant variations because they are closely related to the mean stellar density, approximately scaling as \(\sqrt{M/R^3}\). Since the radius changes only mildly with anisotropy, the dominant contribution to these variations arises mainly from the changes in the stellar mass.

This behavior is physically expected, since the introduction of anisotropy modifies the internal pressure distribution, thereby affecting the equilibrium structure and global stellar properties. On the other hand, the compactness, defined as the ratio between mass and radius, exhibits comparatively smaller variations due to the weaker dependence of the radius on anisotropy. These results therefore suggest that anisotropy primarily impacts the stellar mass, while the remaining quantities respond consistently through their dependence on \(M\) and \(R\). This feature may be relevant when using oscillation modes and gravitational-wave observations as probes of the internal structure of NSs.

Although Table \ref{table} illustrates the influence of anisotropy on representative stable configurations, it is also important to investigate how anisotropy affects the limiting equilibrium models associated with the maximum-mass configurations shown in Fig.~\ref{fig:Mrho_MR}. For this reason, in Table \ref{table3} we report the physical properties of the maximum-mass configurations for the selected values of the anisotropy parameter \(\alpha\). The results presented in Table \ref{table3} confirm the same general trend: the stellar mass is the quantity most strongly affected by anisotropy, whereas the radius remains comparatively stable. Consequently, the variations observed in the oscillation frequencies and compactness mainly reflect their dependence on the stellar mass.

\subsection{On the detectability of the fundamental mode signal}
%%%%%%%%%%%%%%%%%%%%%

%A core-collapse supernova (CCSN) is a cataclysmic explosion that marks the death of a massive star with a mass exceeding approximately $\sim 8 M_{\odot}$. When the stellar core exceeds the effective Chandrasekhar mass $\sim 1.4 M_{\odot}$, it becomes gravitationally unstable and collapses until nuclear densities are reached. This rapid collapse launches a shock wave that starts the CCSN explosion. Gravitational waves and neutrinos are emitted directly from the collapsing core, offering a unique and potentially transformative probe into the physical processes that drive CCSN explosions.  After the supernova explosion, a hot proto-neutron star can be formed, and during the first seconds of its evolution, the gravitational waves are dominated by the fundamental mode $f$ as well as the first pressure mode $p_1$ \cite{Powell_Lasky_2025}. As is well known, a typical core-collapse supernova releases a total energy of approximately $\sim 10^{53}$ erg (0.056 $M_{\odot}$), and we will discuss that some part of the energy could be emitted through the $f$-mode oscillations \cite{Powell_Lasky_2025,Apostolatos2001}.

A core-collapse supernova (CCSN) is a violent explosion marking the end of the life of a massive star with $M \gtrsim 8 M_{\odot}$. Once the stellar core exceeds the effective Chandrasekhar limit ($\sim 1.4 M_{\odot}$), it becomes gravitationally unstable and collapses to nuclear densities. The ensuing collapse drives a shock wave that triggers the CCSN explosion. Gravitational waves and neutrinos are emitted directly from the collapsing core, providing unique probes of the physical mechanisms underlying CCSN dynamics. Following the explosion, a hot proto-neutron star may form; during its early evolution, the gravitational-wave spectrum is dominated by the fundamental $f$-mode and the first pressure mode $p_1$ \cite{Powell_Lasky_2025}. A typical CCSN releases a total energy of $\sim 10^{53}$ erg ($\sim 0.056 M_{\odot}$), and part of this energy may be radiated through $f$-mode oscillations \cite{Powell_Lasky_2025,Apostolatos2001}.

Let us consider the detection of gravitational waves associated with the fundamental oscillation mode of a magnetar formed after a supernova explosion. It is well established that when a gravitational wave reaches the detector, the signal takes the following form:
\begin{equation}
h(t) = he^{-t/\tau}{\rm \sin}[2\pi f t],
\end{equation}
where $f$ is the fundamental mode frequency, $\tau$ is the damping time and the gravitational-wave amplitude $h$ is given by
\begin{equation}
    h \sim A \left( \frac{E_{gw}}{10^{-6}M_{\odot}}    \right)^{1/2}
    \left( \frac{10 {\rm kpc} }{d} \right)
    \left( \frac{1 {\rm kHz}}{f} \right)
    \left( \frac{{1 \rm ms}}{\tau}  \right)^{1/2},
    \label{detection1}
\end{equation}
where $A= 2.4 \times 10^{-20}$, $E_{gw}$ is the energy released through the fundamental mode and $d$ is the distance to the source \cite{Echevarria1989,Apostolatos2001}.

The signal-to-noise ratio at the detector reads \cite{Echevarria1989,Apostolatos2001}
\begin{equation}
\left( \frac{S}{N} \right)^2 = \frac{4 Q_{\textit{F}}^2}{1+4 Q_{\textit{F}}^2} \; \frac{ h^2 \tau}{2 S_n}.
\label{detection2}
\end{equation}
Here, $Q_{\textit{F}} \equiv \pi f \tau$ is the quality factor and $S_n$ is the noise power spectral density.

{From Eqs. \eqref{detection1} and \eqref{detection2}, the energy radiated by the fundamental mode is given by}
\begin{equation}
 \left( \frac{E_{\rm gw}}{M_{\odot}}  \right)  = B C \left( \frac{S}{N} \right)^2 \left( \frac{d}{10 {\rm kpc} }  \right)^2 \left( \frac{f}{1 {\rm kHz}}  \right)^2 \left( \frac{S_n}{1 {\rm Hz^{-1}} } \right) ,
\label{detection3}
\end{equation}
where $B = 3.47 \times 10^{36} $ and $ C = (1+4 Q_{\textit{F}}^2)/4 Q_{\textit{F}}^2$. {In the following, we estimate the minimum energy that must be emitted in order to achieve a signal-to-noise ratio (S/N) greater than $5$, for a source that could be observed within our galaxy, i.e., {at} $d \sim 10\,[\rm kpc]$}.

\begin{table}[tbp]
\centering
%\begin{ruledtabular}
\begin{tabular}{ccccc}
\hline
$\rho_c\,[\rm MeV/fm^3]$&$\alpha$ & $\tau\,[\rm ms]$  & $E^{LV}_{gw}\,[\rm 10^{-7}M_{\odot}]$& $ E^{ET}_{gw}\,[\rm 10^{-10}M_{\odot}]$ \\
\hline
      & $-1.0$ & $445.450$ & $0.8582$  & $2.1454$\\
      & $-0.5$ & $419.308$ & $0.8775$  & $2.2194$\\
$400$ & $0.0$  & $394.512$ & $0.8974$  & $2.2435$\\
      & $+0.5$ & $371.023$ & $0.9183$  & $2.2958$\\ 
      & $+1.0$ & $348.620$ & $0.9399$  & $2.3498$\\
\hline
      & $-1.0$ & $225.026$ & $1.2347$  & $3.0868$\\
      & $-0.5$ & $192.603$ & $1.3079$  & $3.2698$\\
$600$ & $0.0$  & $164.723$ & $1.3865$  & $3.4663$\\
      & $+0.5$ & $140.722$ & $1.4705$  & $3.3676$\\ 
      & $+1.0$ & $120.055$ & $1.5576$  & $3.8941$\\
\hline
\end{tabular}
\caption{\label{table2}{The damping time $\tau$ for NSs, the energy required to excite the fundamental mode, $E^{LV}_{gw}$ and $E^{ET}_{gw}$, for the signal to be detectable by Advanced LIGO-Virgo and the Einstein Telescope, respectively. For this result, we considered the source at $ d \sim 10 $ $\rm [\rm kpc]$, a signal-to-noise ratio S/N $ \sim 5$, two different central densities, and different values of $\alpha$.} } 
%\end{ruledtabular}
\end{table}

{In Table \ref{table2}, we show the minimum energy required for the fundamental mode to be detected}. For this objective, we consider two detectors; the first one is Advanced LIGO-Virgo with {a sensitivity of} $S_n^{1/2} \sim 2 \times 10^{-23}\, \mathrm{[\rm Hz]}^{-1/2}$ at $\sim[\rm kHz]$ \cite{abbott2017_4}, and the second one is the Einstein Telescope with {a sensitivity} $S_n^{1/2} \sim 10^{-24} \, \mathrm{[\rm Hz]}^{-1/2}$ at {a similar frequency band} \cite{2017CQGra..34d4001A}. 

%{\bf Our results show that for stars of about $1.4\,M_{\odot}$ the $f$-mode frequency is close to $2.0$ [kHz] and the damping time is about $164.0$ [ms]. Also, as can be seen, at a distance of $10$ [kpc], the minimum energy emitted in the GW must be $1.53 \times10^{-7} M_{\odot}$. As can be seen, this required energy is much lower than the energy released in a CCSN explosion, which is about ($10^{-5} - 10^{-6} M_{\odot}$ ).}

For a fixed central energy density $\rho_c=600\,[\rm MeV/fm^3]$, varying the anisotropy from $\alpha=-1.0$ to $\alpha=+1.0$ produces measurable shifts in the $f$-mode properties and the GW energy needed for detectability. For $\alpha=-1.0$ we obtain $M=1.5498\,M_\odot$, $f=1.8863\,[\rm kHz]$, $\tau=225.026\,[\rm ms]$, with detection thresholds $E^{LV}_{\mathrm{gw}}>1.2347\times10^{-7}M_\odot$ (Advanced LIGO/Virgo, S/N$>5$) and $E^{ET}_{\mathrm{gw}}>3.0868\times10^{-10}M_\odot$ (Einstein Telescope, S/N$>5$). For $\alpha=+1.0$ we find $M=1.6902\,M_\odot$, $f=2.1187\,[\rm kHz]$, $\tau=120.055\,[\rm ms]$, and thresholds $E^{LV}_{\mathrm{gw}}>1.5576\times10^{-7}M_\odot$ and $E^{ET}_{\mathrm{gw}}>3.8941\times10^{-10}M_\odot$. This systematic dependence of the $f$-mode properties on the anisotropy parameter suggests that anisotropy may leave observable imprints on the gravitational-wave signal. However, given current detector sensitivities in the $[\rm kHz]$ band and additional modelling uncertainties, such effects should be interpreted with caution. Future high-precision observations, combined with improved theoretical modelling, may help to place constraints on the degree of anisotropy in NSs.

\subsection{Tidal deformability}
\label{4tidaldef}
{
\begin{figure}[h]
    \centering
    \includegraphics[width=0.49\textwidth]{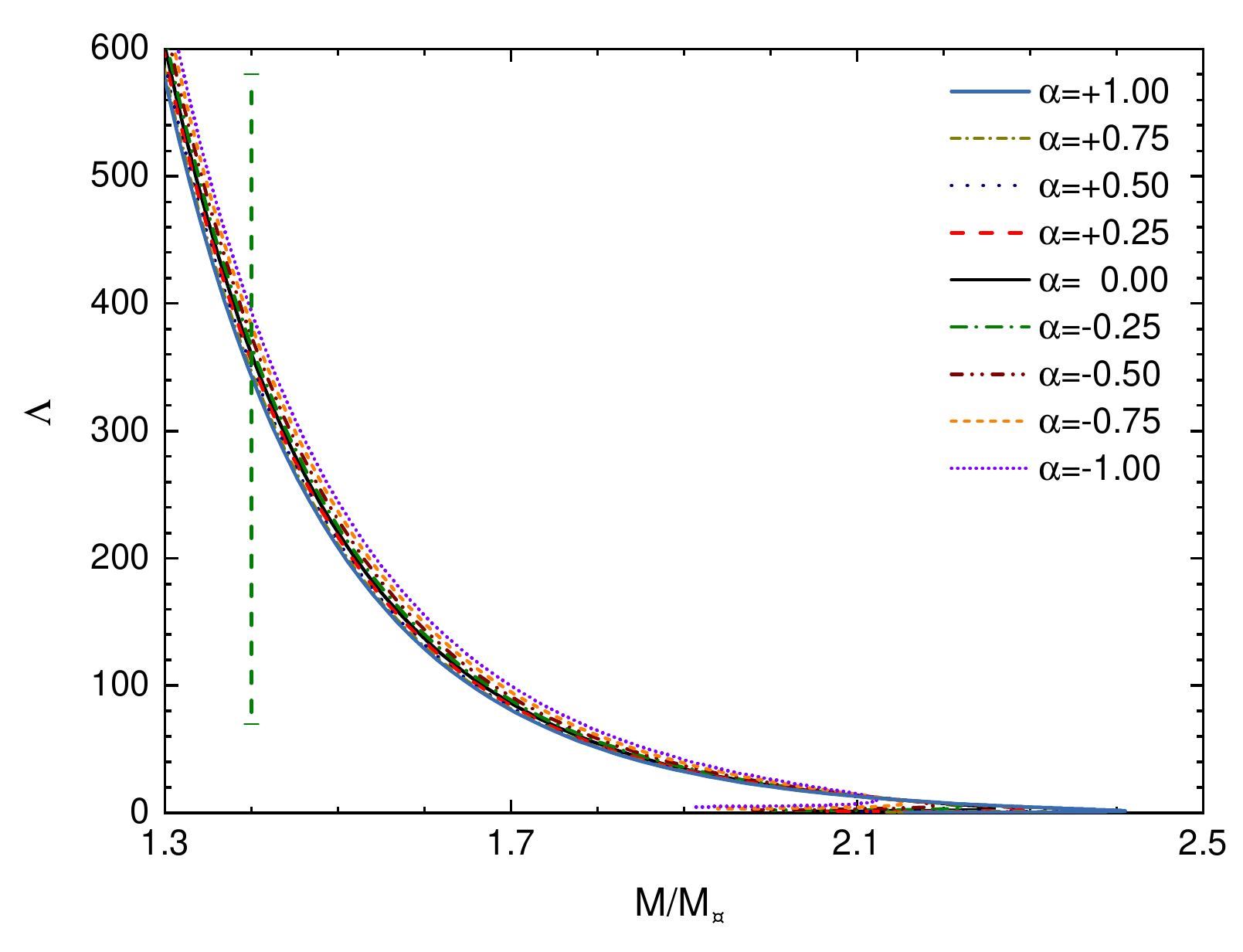}
    \includegraphics[width=0.49\textwidth]{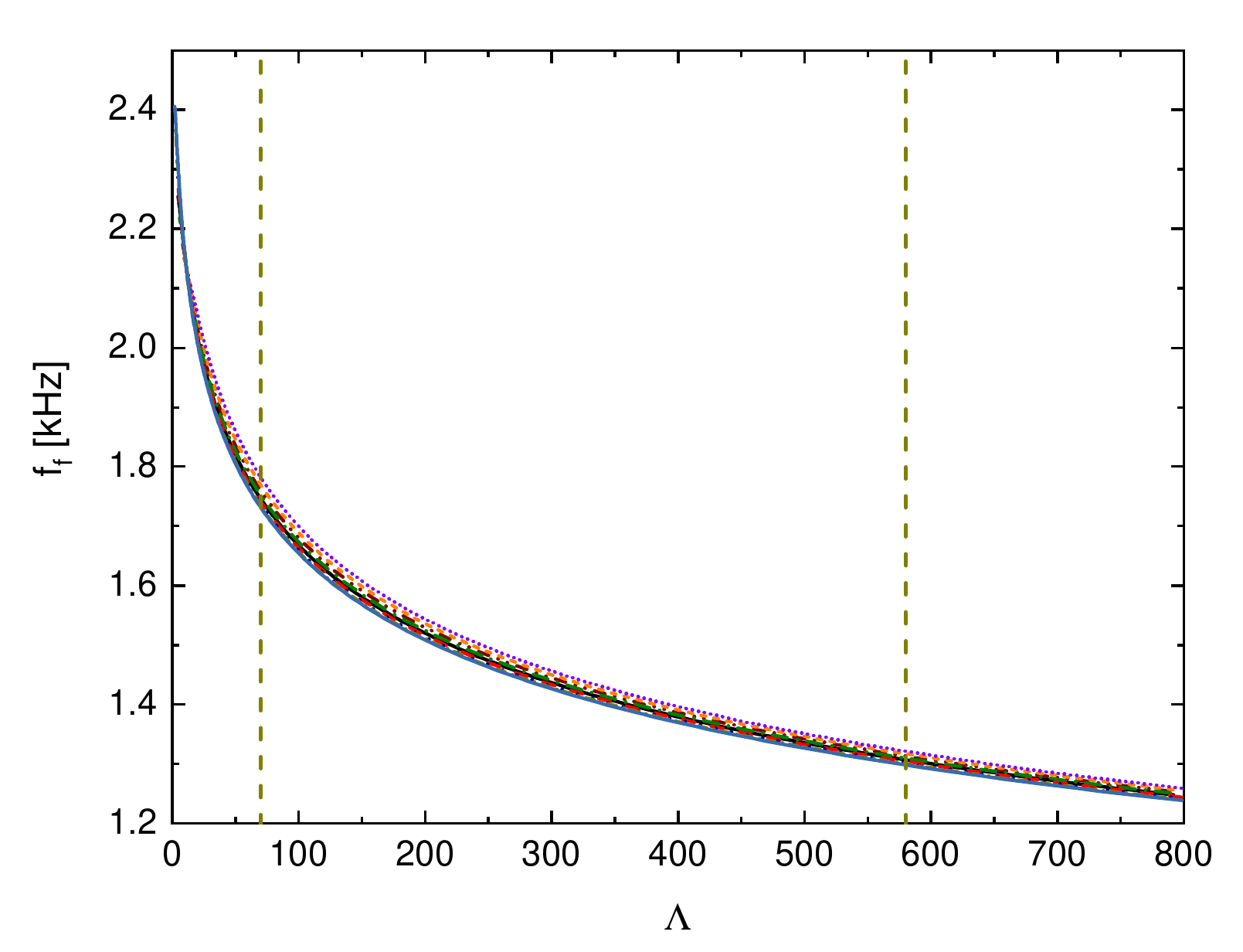}
    \caption{Left: Dimensionless tidal deformability against the total mass for different values of $\alpha$. Right: Oscillation frequency $f_f$ as a function of the dimensionless tidal deformability for nine values of $\alpha$. In both panels, the vertical dashed line indicates the dimensionless tidal deformability constraint $\Lambda_{1.4}=190^{+390}_{-120}$ from the event GW$170817$ reported by the LVC in Ref. \cite{2018PhRvL.121p1101A}.}
    \label{fig:lvsM}
\end{figure}

The use of GW$170817$ tidal-deformability constraints to test dense-matter models has been explored mainly to place limits on the properties of compact stars within the corresponding theoretical scenarios. For instance, \cite{lourenco2021} investigated this parameter by considering quark matter in the color-flavor-locked phase of color superconductivity, \cite{li2021} examined the role of isospin effects in strange quark matter, \cite{xu2022} analyzed the tidal deformability assuming a quasiparticle model that incorporates nonperturbative aspects of quantum chromodynamics in the low-density regime, and \cite{2018PhRvC..98c5804M} used the tidal deformability of NSs to constrain the nuclear-matter EOS. In a context closely related to the present work, Biswas and Bose~\cite{Biswas2019} investigated the impact of pressure anisotropy on the tidal deformability of compact stars and concluded that anisotropy modifies this parameter, while the results remain compatible with GW$170817$ within observational uncertainties. In this context, our analysis complements previous work by studying, for the same anisotropic stellar sequences, both the tidal deformability and its correlation with the $f$-mode frequency in a complete general-relativistic framework.

Fig. \ref{fig:lvsM} plots, in its left and right panels respectively, the dimensionless tidal deformability $\Lambda$ against the total mass $M/M_{\odot}$ and the oscillation frequencies of the fundamental mode $f_f$ as a function of the dimensionless tidal deformability $\Lambda$ for several choices of the anisotropy parameter $\alpha$. These curves are compared with the observational range for $\Lambda_{1.4} = 190^{+390}_{-120}$ reported by LIGO-Virgo \cite{2018PhRvL.121p1101A}. In the left panel, the tidal deformability decreases monotonically as the total mass increases, up to the maximum-mass configuration. After this point, the curves bend clockwise, and the tidal deformability continues to decrease as the total mass diminishes. Additionally, the influence of anisotropy on the deformability response is clearly seen: when $\alpha$ is positive, the calculated $\Lambda$ values are lower for a given mass, whereas negative $\alpha$ leads to larger values. Importantly, all curves remain within the $\Lambda_{1.4}$ band determined by LIGO-Virgo. In the right panel, the curves show that, as the tidal deformability increases, the $f$-mode frequency decreases monotonically for all values of the anisotropy parameter. This trend is consistent with the approximate scalings \(f\sim\sqrt{M/R^3}\) and \(\Lambda\sim(R/M)^5\) \cite{Wen2019}, according to which less compact stars have larger tidal deformabilities and lower oscillation frequencies.

\begin{figure}[h]
    \centering
    \includegraphics[width=0.49\textwidth]{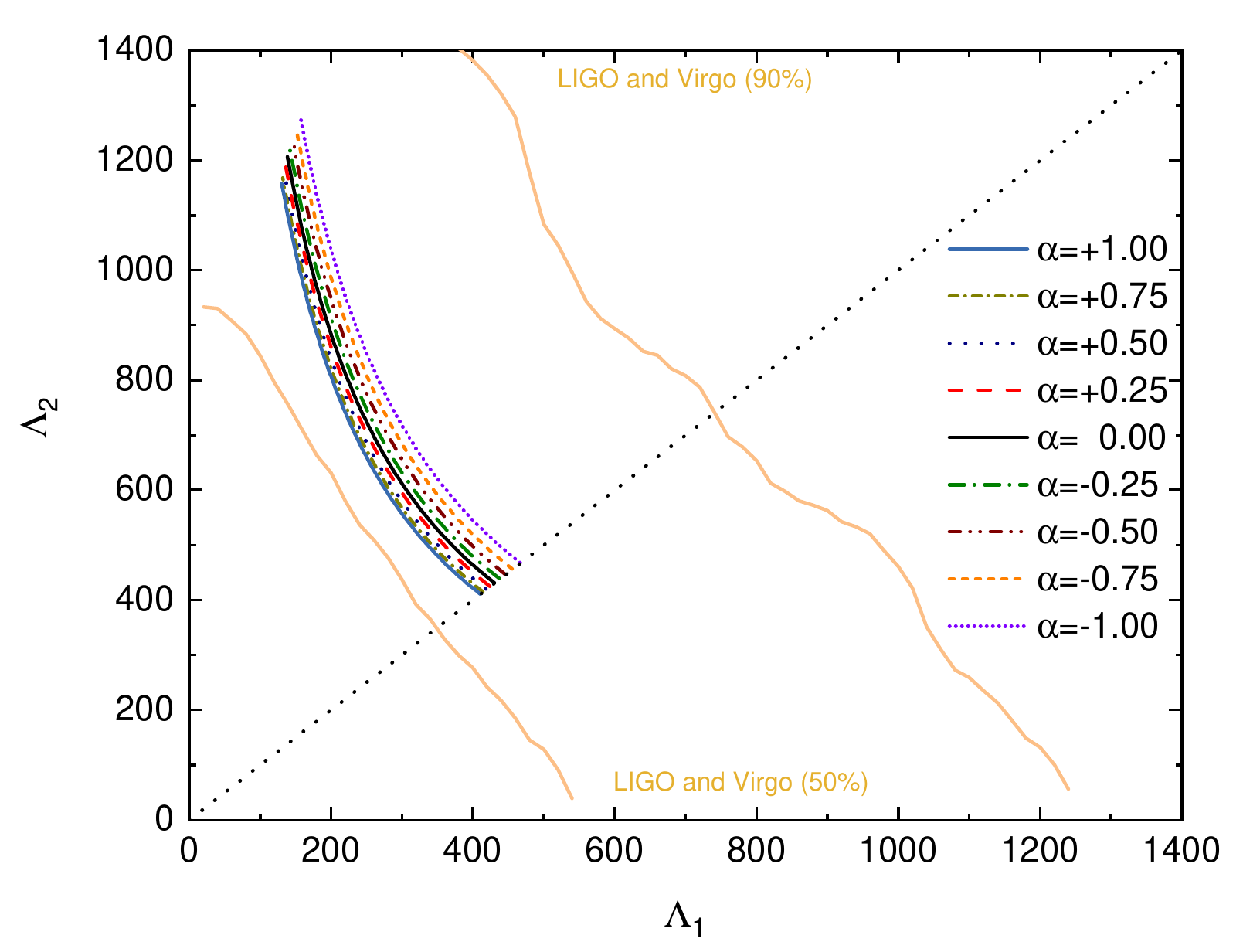}
    \caption{The dimensionless tidal deformabilities of the GW$170817$ event components are shown for different values of the anisotropic parameter $\alpha$. The yellow contours show the 50\% and 90\% credible regions reported by the LIGO--Virgo Collaboration for GW170817 \cite{2018PhRvL.121p1101A}, while the dotted diagonal line marks the points where $\Lambda_1 = \Lambda_2$.}
    \label{fig:l1l2}
\end{figure}

Using the observational data provided by LVC, the study in \cite{2018PhRvL.121p1101A}   established bounds on $\Lambda_1$ and $\Lambda_2$, which quantify the dimensionless tidal deformability of the binary system. Here, $\Lambda_1$ corresponds to the tidal deformability of the more massive star, while $\Lambda_2$ refers to that of its companion. Fig. \ref{fig:l1l2} presents the $\Lambda_1$--$\Lambda_2$ plane, where each curve is generated by selecting a specific value of $M_1$ and then computing $M_2$ using the chirp mass $\mathcal{M} = 1.188\,M_\odot$ \cite{abbott2017_4}, defined as
\begin{equation}
\mathcal{M} = \frac{(M_1 M_2)^{3/5}}{(M_1+M_2)^{1/5}}.
\end{equation}
The stellar masses considered fall within the intervals $1.36 \leq M_1/M_\odot \leq 1.60$ and $1.17 \leq M_2/M_\odot \leq 1.36$, respectively. In addition, the $50\%$ and $90\%$ credibility contours for the GW$170817$ event, provided by LVC under the low-spin assumption, are included in the plot. As $\alpha$ varies, a clear systematic trend is observed: negative values of $\alpha$ shift the curves toward higher tidal deformabilities, whereas positive values drive them toward smaller values. This behavior highlights the sensitivity of the tidal deformability to the anisotropic pressure distribution inside the star.\label{4.4.3}

%For example, for $\rho_c=600\,[\rm MeV/fm^3]$, when $\alpha=-1.0$, for a star of mass of $1.1918\,M_\odot$, with $f=2.1073\,[\rm kHz]$ and $\tau=187.033\,[\rm ms]$, the required energy is $E^{LV}_{gw}>1.5409 \times10^{-7}M_{\odot}$ for a (S/N)$>5$ at Advanced LIGO/VIRGO and $E^{ET}_{gw}>1.5409\times10^{-10}M_{\odot}$ for a (S/N)$>5$ at the Einstein Telescope. Moreover, when $\alpha=+1.0$, for a star with a mass of $1.78\,M_\odot$, $f=2.0781\,[\rm kHz]$ and $\tau=153.927\,[\rm ms]$, then we can see that $E^{LV}_{gw}>1.4985 \times  10^{-7}M_{\odot}$ for a (S/N)$>5$ at Advanced LIGO/VIRGO and $E^{ET}_{gw}>3.7462\times 10^{-10}M_{\odot}$ for a (S/N)$>5$ at the Einstein Telescope. From these results, we can see that anisotropy analysis can also play an important role in determining the amount of energy needed to excitep the fundamental mode.

}

%%%%%%%%%%%%%%%%%%%%%
\section{Conclusions}\label{conclusions}
%%%%%%%%%%%%%%%%%%%%%

This article investigates the effects of anisotropy on both the stellar structure and the $f$-mode oscillations of NSs. To this end, we used the stellar equilibrium equations, including the anisotropic factor, and derived the nonradial oscillation equations involving such a factor within the fully general relativity context. For the fluid contained in the compact star, we assume that the stellar matter obeys an EOS derived from a consistent matching between microscopic nuclear theory and pQCD calculations, joined through a piecewise polytropic interpolation scheme, and the anisotropy is described by the profile $\sigma=\alpha p_r(1-\frac{1}{g_{11}})^2$ (see Ref.~\cite{2024PhRvD.110h3020L,2026PhRvD.113d3025G,2026arXiv260106962Y}). 

Regarding the equilibrium configurations, we found that the NS structure is affected by the presence of anisotropic pressure. For certain ranges of central energy density, we observed both larger (smaller) mass and radius values when using higher (lower) values of $\alpha$. In addition, variations in the anisotropy parameter $\alpha$ alter the internal pressure distribution of the star, thereby influencing the resulting mass-radius configurations. As shown in Fig.~\ref{fig:Mrho_MR}, this leads to stellar models that are in better agreement with current observational constraints, including NICER mass-radius measurements and heavy pulsar masses. This indicates that anisotropy can enhance the phenomenological viability of the EOS adopted in this work. This suggests that anisotropy may play a relevant role by providing complementary information on the internal structure and composition of NSs.

In our study of nonradial oscillation modes, we found that the $f$-modes are systematically affected by anisotropy, with relative variations depending on the stellar configuration. According to Fig.~\ref{fig:fmodes_OmegaM}, for specific ranges of total mass, we observed that the $f$-modes increase (or decrease) as the anisotropy parameter $\alpha$ increases (or decreases). It is worth noting that the equations governing nonradial oscillations depend on the chosen anisotropic profile; in this case, they are determined by the form of equation \eqref{anisotropic_eos}.

We also investigated the impact of anisotropy on the detectability of the fundamental oscillation mode. Our results indicate that the $f$-mode signal from NSs with masses of $\sim 1.7\,M_\odot$ and an anisotropy parameter $\alpha \sim 1.0$ could be detected within our galaxy with future GW detectors, i.e., at distances of approximately $\sim 10\,[\rm kpc]$.

We have examined the consistency between the dimensionless tidal deformability of anisotropic NSs and the observational constraints reported by the LVC for the GW$170817$ event. The results presented in this work fall within the observational bounds provided by the LIGO–Virgo Collaboration. We found that, when $\alpha > 0$, the tidal deformability decreases for increasing $\alpha$, whereas for $\alpha < 0$, it increases as $\alpha$ becomes more negative. This behavior suggests that anisotropy could mimic the tidal response of different isotropic configurations, thereby potentially introducing a degeneracy between the anisotropy parameter and the underlying EOS. As a result, current observational constraints may not be sufficient to uniquely disentangle anisotropic effects from other physical uncertainties. Future high-precision gravitational-wave observations, together with improved modelling, will be essential to disentangle anisotropic effects from other physical uncertainties and to fully exploit the potential of NS asteroseismology.

While we have shown that anisotropic pressure affects the equilibrium and nonradial oscillation spectrum of non-rotating NSs within a complete general relativistic context, in rotating configurations it couples nontrivially to centrifugal effects, thereby modifying the stellar deformation, as characterized by the difference between the equatorial and polar radii. In particular, increasing the anisotropy parameter in the Bowers-Liang model \cite{bowers1974} leads to more massive and compact configurations that become progressively less susceptible to rotational deformation, as reflected in the reduction of the eccentricity and the convergence between the equatorial and polar radii \cite{Pattersons2021}. Motivated by these results, in the future we plan to investigate the impact of anisotropic pressure on the deformation of rotating NSs within the framework of our EOS model.

Although anisotropy produces systematic variations in the $f$-mode frequency, the magnitude of these effects is relatively moderate (of the order of a few percent) and may be comparable to uncertainties arising from the EOS and other modelling assumptions. Nevertheless, the clear and systematic dependence of the $f$-mode properties on the anisotropy parameter found in this work suggests that anisotropy may leave identifiable imprints on the gravitational-wave signal. However, given current detector sensitivities in the $[\rm kHz]$ regime, where the $f$-mode signal is expected, the extraction of such effects from observational data remains challenging \cite{Apostolatos2001,Echevarria1989,2017CQGra..34d4001A}. Future improvements in detector sensitivity and theoretical modelling will be essential to disentangle anisotropic effects from other sources of uncertainty and to place meaningful constraints on its magnitude.

The novelty of this work lies in the combined analysis of $f$-mode oscillations and tidal deformability within a consistent anisotropic framework in full general relativity. In contrast to previous studies, which typically focus on either oscillation modes or tidal properties separately, our results show that anisotropy induces correlated modifications in both quantities. This provides a more comprehensive picture of how anisotropic effects may manifest in GW observations.

%%%%%%%%%%%%%%%%%%%%%%%%%%%%%%%%%%%%%%%%%%%%%

\acknowledgments
The authors thank Shu Yan Lau, Siddarth Ajith, Victor Guedes, and Kent Yagi for useful discussions regarding the work. JDVA thanks Universidad Privada del Norte and Universidad Nacional Mayor de San Marcos for the financial support - RR No.$\,005753$-$2021$-R$/$UNMSM under the project number B$21131781$. GOC is thankful to CNPq (grant No. 176956/2025-5). JMZP acknowledges the financial support provided by FAPERJ under Process No.~SEI-260003/000308/2024. CVF acknowledges the financial support of the productivity program of the Conselho Nacional de Desenvolvimento Cient\'ifico e Tecnol\'ogico (CNPq), with Project No.~304569/2022-4. CHL is thankful to the S\~ao Paulo Research Foundation FAPESP (Grant No.~2020/05238-9) and to CNPq (Grants No.~401565/2023-8 and 305327/2023-2).
%%%%%%%%%%%%%%%%%%%%%%%%%%%%%%%%%%%%%%

\appendix

\section{Nonradial perturbations of relativistic spherically symmetric stars in the presence of an anisotropic fluid}\label{Appendix_A}

\subsection{Perturbative variables}

For even-parity harmonics, the equations of motion are found through the perturbed Einstein's field equation \eqref{perturbed_FE}, by employing the metric perturbation \eqref{metric_perturbation} and the Lagrangian fluid displacement vector components in the form
\begin{eqnarray}
&&\xi^r=\frac{1}{r^2}e^{-\Lambda}WY^{\ell m},\label{fluid_displacement_r}\\
&&\xi^{\theta}=-\frac{1}{r^2}V\partial_{\theta}Y_{\ell m},\label{fluid_displacement_theta}\\
&&\xi^{\phi}=-\frac{1}{r^2\sin^2\theta}V\partial_{\phi}Y_{\ell m},\label{fluid_displacement_phi}
\end{eqnarray}
where $W$ and $V$ are functions that depend on $t$ and $r$. Moreover, the non-perturbed four-velocity and radial unit vector are used, which are respectively placed as
\begin{eqnarray}\label{u_k_0}
&&u^{(0)\mu}=\left(\frac{1}{e^{\Psi}},0,0,0\right)\quad {\rm and}\quad k^{(0)\mu}=\left(0,\frac{1}{e^{\Lambda}},0,0\right), 
\end{eqnarray}
respectively. The perturbed four-velocity components are set in the form
\begin{eqnarray}\label{u_varios}
&&\delta u^t=\frac{1}{2}\frac{H Y_{\ell m}}{e^{\Psi}},\\
&&\delta u^{r}=\frac{1}{e^{\Psi}}\frac{d\xi^r}{dr}=\frac{\dot{W}Y_{\ell m}}{r^2e^{\Psi+\Lambda}},\\
&&\delta u^{\theta}=\frac{1}{e^{\Psi}}\frac{d\xi^{\theta}}{dr}=-\frac{\dot{V}\partial_{\theta}Y_{\ell m}}{r^2e^{\Psi}},\\
&&\delta u^{\phi}=\frac{1}{e^{\Psi}}\frac{d\xi^{\phi}}{dr}=-\frac{\dot{V}\partial_{\phi}Y_{\ell m}}{r^2e^{\Psi}\sin^2\theta},
\end{eqnarray}
and the perturbed radial unit vector components are given by
\begin{eqnarray}\label{k_varios}
&&\delta k^0=\frac{1}{e^{\Psi}}\left[\frac{H_1}{e^{\Psi+\Lambda}}+\frac{\dot{W}}{r^2e^{\Psi}}\right]Y_{\ell m}\quad {\rm and}\quad\delta k^1=-\frac{1}{2}\frac{H}{e^{\Lambda}}Y_{\ell m}.
\end{eqnarray}
The relations defined in Eqs.~\eqref{u_k_0}, \eqref{u_varios}, and \eqref{k_varios} satisfy the equalities $g_{\mu\nu}k^{\mu}k^{\nu}=1$, $g_{\mu\nu}u^{\mu}k^{\nu}=0$, and $g_{\mu\nu}u^{\mu}u^{\nu}=-1$ in their both non-perturbed and perturbed form.

\subsection{Perturbation equations}

To obtain the equations of motion of the pulsating configuration, following Ref.~\cite{thorne1967}, the Lagrangian perturbation of the baryon number density $N$ is
\begin{equation}\label{DeltaN_N}
    \frac{\Delta N}{N}=\xi^{k}_{\,\,|k}-\frac{1}{2}\frac{\delta\left[^{(3)}g\right]}{^{(3)}g}.
\end{equation}
The term $\xi^{k}_{\,\,|k}$ denotes the covariant derivative in the $3$-dimensional geometry at constant time, and the terms $\delta\left[^{(3)}g\right]$ and $^{(3)}g$ depict the determinant of the perturbed and unperturbed $3$-dimensional metric, respectively. By employing the background metric \eqref{line_element_background}, the metric perturbation \eqref{metric_perturbation}, and the fluid displacement vectors \eqref{fluid_displacement_r}-\eqref{fluid_displacement_phi}, the equation \eqref{DeltaN_N} yields 
\begin{equation}\label{delta_N}
   \frac{\Delta N}{N}=-\frac{1}{2}HY_{\ell m}-\frac{e^{-\Lambda}}{r^2}W'Y_{\ell m}-\frac{V}{r^2}\ell(\ell+1)Y_{\ell m}-KY_{\ell m}.
\end{equation}
Using the baryon conservation equation, $\nabla_{\mu}(Nu^{\mu})=0$, the local law of energy conservation ($u^{\nu}\nabla^{\mu}T_{\mu\nu}$=0) in terms of Lagrangian perturbations yield (see \cite{2024PhRvD.110h3020L}):
\begin{eqnarray}\label{delta_rho}
    &&\Delta\rho=(\rho+p_r)\frac{\Delta N}{N}-\sigma\left(\Delta U^{\theta}_{\theta}+\Delta U^{\phi}_{\phi}\right),\\
    &&\Delta U^{\theta}_{\theta}+\Delta U^{\phi}_{\phi}=\left(K+2e^{-\Lambda}\frac{W}{r^2}+\frac{\ell(\ell+1)}{r^2}V\right)Y_{\ell m}.
\end{eqnarray}
Employing the relations of the thermodynamic functions in their Lagrangian and Eulerian forms, $\Delta\rho\thickapprox\delta\rho+\frac{\partial\rho^0}{\partial r}\xi^r$, and equation \eqref{delta_N}, we find that equation \eqref{delta_rho} takes the form:
\begin{eqnarray}
&&\hspace{-1cm}\delta\rho=-(p_r+\rho)\left[\frac{H}{2}+\frac{e^{-\Lambda}}{r^2}W'+\frac{V}{r^2}\ell(\ell+1)+K\right]Y_{\ell m}-\sigma\left(K+\frac{V}{r^2}\ell(\ell+1)\right.\nonumber\\
&&\left.+\frac{2e^{-\Lambda}}{r^3}W\right)Y_{\ell m}-\rho'\frac{e^{-\Lambda}}{r^2}WY_{\ell m}.\label{delta_rho}
\end{eqnarray}
The perturbed radial pressure can be determined using the EOS $p_r=p_r(\rho)$, where $\delta p_r=\frac{dp_r}{d\rho}\delta\rho$. From Eq. \eqref{perturbed_FE}, the nonzero perturbed Einstein field equations are given by
\begin{eqnarray}
&&\hspace{-0.4cm}H'+\frac{He^{2\Lambda}}{r}\left[4\pi r^2(p_r-\rho)+1+\frac{\ell(\ell+1)}{2}\right]=rK''-\frac{8\pi e^{\Lambda}W}{r}\left[\rho'+\frac{2\sigma}{r}\right]-\frac{8\pi e^{2\Lambda}V\ell(\ell+1)}{r}\left(p_r\right.
\nonumber\\
&&\hspace{-0.4cm}\left.+\rho+\sigma\right)-\frac{8\pi e^{\Lambda}W'}{r}\left(p_r+\rho\right)-\frac{Ke^{2\Lambda}}{r}\left[8\pi r^2(p_r+\rho+\sigma)-1+\frac{\ell(\ell+1)}{2}\right]-e^{2\Lambda}\left[4\pi\rho r^2\right.\nonumber\\
&&\left.+\frac{5m}{r}-3\right]K',\label{var_H_der}\\
&&\hspace{-0.4cm}\ell(\ell+1)H_1=2r^2{\dot K}'+2r{\dot K}e^{2\Lambda}\left(-4\pi r^2p_r-\frac{3m}{r}+1\right)-2r{\dot H}-16\pi(p_r+\rho)e^{\Lambda}{\dot W},\\
&&\hspace{-0.4cm}H_1'=-\frac{2e^{\Lambda}H_1}{r}\left[\frac{m}{r}+2\pi r^2(p_r-\rho)\right]+e^{2\Lambda}{\dot H}+e^{2\Lambda}{\dot K}+16\pi\left(\rho+p_r+\sigma\right)e^{2\Lambda}{\dot V},\\
&&\hspace{-0.4cm}{\ddot K}-e^{2\Psi-2\Lambda}K''-\frac{Ke^{2\Psi}}{r^2}\left(8\pi r^2\left(\Gamma p_r-\rho-p_r-\sigma+\sigma\frac{dp_r}{d\rho}\right)-\ell(\ell+1)+2\right)\nonumber\\
&&\hspace{-0.4cm}-\frac{2K'e^{2\Psi}}{r}\left(1-\frac{m}{r}+2\pi r^2(p_r-\rho)\right)-\frac{e^{2\Psi}H}{r^2}\left(4\pi(p_r+\rho+\Gamma p_r)r^2-2+\frac{4m}{r}\right)\nonumber\\
&&\hspace{-0.4cm}+\frac{W'e^{2\Psi-\Lambda}}{r^2}8\pi\left(\rho+p_r-\Gamma p_r\right)+\frac{Ve^{2\Psi}}{r^2}8\pi\ell(\ell+1)\left(\rho+p_r+\sigma-\Gamma p_r-\sigma\frac{dp_r}{d\rho}\right)\nonumber\\
&&+W\left(\rho'-p_r'+\frac{2\sigma}{r}\left(1-\frac{dp_r}{d\rho}\right)\right)\frac{8\pi e^{2\Psi-\Lambda}}{r^2}=0,\\
&&{\dot H}_1=H'e^{2\Psi}-K'e^{2\Psi}+\frac{He^{2\Psi+2\Lambda}}{r}\left(\frac{2m}{r}+8\pi r^2p_r\right).
\end{eqnarray}

From the perturbed conservation law of the energy-momentum tensor components $\delta\left(\nabla_{\nu}T^{\nu}_1\right)$ and $\delta\left(\nabla_{\nu}T^{\nu}_2\right)$, we respectively obtain: 
\begin{eqnarray}
&&\hspace{-0.4cm}-\frac{H'}{2}(p_r+\rho)e^{\Psi}-e^{\Psi}\Psi'(p_r+\rho)\left[\frac{H}{2}+\frac{V}{r^2}\ell(\ell+1)+K\right]-e^{\Psi}\Psi'\sigma\left[\frac{2e^{-\Lambda}}{r^3}W\right.\nonumber\\
&&\hspace{-0.4cm}\left.+\frac{V}{r^2}\ell(\ell+1)+K\right]-\left\{e^{\Psi}\left(p_r+\rho\right)\frac{dp_r}{d\rho}\left[\frac{H}{2}+\frac{e^{-\Lambda}}{r^2}W'+\frac{V}{r^2}\ell(\ell+1)+K\right]\right.\nonumber\\
&&\left.+e^{\Psi}\sigma\frac{dp_r}{d\rho}\left(K+\frac{\ell(\ell+1)V}{r^2}+\frac{2e^{-\Lambda}W}{r^3}\right)\right\}'-K'\sigma e^{\Psi}+(p_r+\rho)\frac{e^{\Lambda-\Psi}}{r^2}{\ddot W}\nonumber\\
&&\hspace{-0.4cm}+(p_r+\rho)e^{\Psi}\left[H'-K'+\frac{H}{r}e^{2\Lambda}\left(\frac{2m}{r}+8\pi r^2p_r\right)\right]+(p_r+\rho)\left(\Psi'^2+\Psi''-\Psi'\Lambda'\right.\nonumber\\
&&\hspace{-0.4cm}\left.-\frac{2\Psi'}{r}\right)\frac{e^{\Psi-\Lambda}}{r^2}W-\frac{2}{r^3}\left(\frac{3\sigma}{r}-\sigma'-\sigma\left(\frac{2me^{2\Lambda}}{r^2}+4\pi e^{2\Lambda}(p_r-\rho)r\right)\right)e^{\Psi-\Lambda}W-\frac{2\sigma}{r^3}e^{\Psi-\Lambda}W'\nonumber\\
&&+p_r'\Psi'\frac{e^{\Psi-\Lambda}}{r^2}W-\frac{2e^{\Psi}}{r}(\delta{\tilde\sigma})=0,\label{delta_T1nu}\\
&&\hspace{-0.4cm}(p_r+\rho+\sigma)\frac{\ddot{V}}{e^{2\Psi}}+\frac{H}{2}\left(p_r+\rho+\Gamma p_r\right)+\frac{W'}{r^2e^{\Lambda}}\Gamma p_r+\frac{V\ell(\ell+1)}{r^2}\left(\Gamma p_r\right.\nonumber\\
&&\hspace{-0.4cm}\left.+\sigma\frac{dp_r}{d\rho}\right)+K\left(\Gamma p_r+\sigma\frac{dp_r}{d\rho}\right)+W\left(\frac{p_r'}{r^2e^{\Lambda}}+\frac{2\sigma}{e^{\Lambda}r^3}\frac{dp_r}{d\rho}\right)-\delta{\tilde\sigma}=0,\label{delta_T2nu}
\end{eqnarray}
where $\Gamma=(p_r+\rho)\frac{dp_r}{d\rho}$ and $\delta\sigma=\delta{\tilde\sigma}Y_{\ell m}$. %All equations presented in this section agree with those reported in \cite{sotani_thesis} for the isotropic case; $\sigma=0$ ($p_r=p$) and $\delta{\tilde\sigma}=0$. 
It is important to mention that the term $\delta{\tilde\sigma}$ is renamed as $\delta{\sigma}$. In addition, for the anisotropic profile $\sigma=\sigma(p_r,g_{11})$, we found that
\begin{equation}\label{delta_sigma_a}
\delta\sigma=\frac{\partial\sigma}{\partial p_r}\delta p_r+\frac{\partial\sigma}{\partial g_{11}}h_{11}.
\end{equation}
This last relation is also used to determine Eqs. \eqref{eq_X_tilde}-\eqref{eq_h}. { It is important to mention that, in the Cowling approximation, $h_{\mu\nu}$=0, Eqs. \eqref{delta_rho}-\eqref{delta_sigma_a} reproduce the nonradial oscillation equations reported in Ref. \cite{doneva2012}.}

\section{Perturbation functions near \lowercase{$r=0$}}\label{Appendix_B}

For the isotropic case, $\sigma=0$, the spacetime perturbation and fluid functions expanded in a power series near $r=0$ have been previously reported in Ref. \cite{detweiler1983}. This appendix shows these expanded functions for the anisotropic case, $\sigma\neq0$. Since we seek solutions to the perturbation equations \eqref{eq_H1_tilde_der}--\eqref{eq_X_tilde_der},  the perturbation functions ${\tilde H}_1(r)$, ${\tilde K}(r)$, ${\tilde W}(r)$, and ${\tilde X}(r)$ near the center of the star are considered in the form 
\begin{eqnarray}
&&{\tilde H}_1(r)={\tilde H}_1(0)+\frac{1}{2}{\tilde H}_1''(0)r^2+O(r^4),\\
&&{\tilde K}(r)={\tilde K}(0)+\frac{1}{2}{\tilde K}''(0)r^2+O(r^4),\\
&&{\tilde W}(r)={\tilde W}(0)+\frac{1}{2}{\tilde W}''(0)r^2+O(r^4),\\
&&{\tilde X}(r)={\tilde X}(0)+\frac{1}{2}{\tilde X}''(0)r^2+O(r^4),
\end{eqnarray}
where the first-order -${\tilde H}_1(0)$, ${\tilde K}(0)$, ${\tilde W}(0)$, and ${\tilde X}(0)$- and second-order coefficients -${\tilde H}_1''(0)$, ${\tilde K}''(0)$, ${\tilde W}''(0)$, and ${\tilde X}''(0)$- are constants. The first-order terms are given by:
\begin{eqnarray}
&&\hspace{-0.4cm}{\tilde H}(0)={\tilde K}(0),\\
&&\hspace{-0.4cm}{\tilde X}(0) = (\rho_{_0}+p_{r{_0}})e^{\Psi_0}\left[-\frac{\omega^2e^{-2\Psi_{_0}}}{\ell}{\tilde W}(0)+\frac{4\pi}{3}\left(\rho_{_0}+3p_{r_0}\right){\tilde W}(0)-\frac{{\tilde K}(0)}{2}\right],\label{B.6}\\
&&\hspace{-0.4cm}{\tilde H}_1{(0)} = \left[\frac{2\ell {\tilde K}(0)-16\pi(\rho_0+p_{r_0}){\tilde W}(0)}{\ell(\ell+1)}\right],\\
&&\hspace{-0.4cm}{\tilde V}(0)=-\frac{1}{\ell}{\tilde W}(0),
\end{eqnarray}
with the constants $\rho_{_0}$, $p_r{_0}$, and $\Psi_{_0}$ representing the first-order terms of the power-series expansions:
\begin{eqnarray}
&&\rho(r)=\rho_{_0}+\frac{1}{2}\rho_{_2}r^2,\label{rho_exp}\\
&&p_r(r)=p_{r_0}+\frac{1}{2}p_{r_2}r^2+\frac{1}{4}p_{r_4}r^4,\label{p_exp}\\
&&\Psi(r)=\Psi_{_0}+\frac{1}{2}\Psi_{_2}r^2+\frac{1}{4}\Psi_{_4}r^4,\label{phi_exp}\\
&&\sigma(r)=\sigma_{_0}+\frac{1}{2}\sigma_{_2}r^2+\frac{1}{4}\sigma_{_4}r^4,\label{sigma_exp}
\end{eqnarray}
and $\alpha$ being the dimensionless anisotropic constant. Through equation \eqref{anisotropic_eos}, we can see that in the star's center, because $g_{11}=1$ (see Eq. \eqref{eq_phi}), we have $\sigma_{_0}=0$. The second and fourth-order coefficients of the power-series \eqref{rho_exp}-\eqref{sigma_exp} are
\begin{eqnarray}
&&\hspace{-0.4cm}\Psi_{_2}= \frac{4\pi}{3}(\rho_{_0}+3p_{r_0}),\label{phi2_exp}\\
&&\hspace{-0.4cm}\rho_{_2}=-\frac{(\rho_{_0}+p_{r_0})^2}{\Gamma\,p_{r_0}}\Psi_{_2},\label{rho2_exp}\\
&&\hspace{-0.4cm}\sigma_{_2}=0,\label{sigma2_exp}\\
&&\hspace{-0.4cm}p_{r_2}=-(\rho_{_0}+p_{r_0})\Psi_2,\label{p2_exp}\\
&&\hspace{-0.4cm} \Psi_{_4} = \frac{2\pi}{5}(\rho_{_2}+5p_{r_2})+\frac{32\pi^2}{9}\rho_{_0}(\rho_{_0}+3p_{r_0}),\label{phi4_exp}\\
&&\hspace{-0.4cm}\sigma_{_4} = \left(\frac{16\pi\rho_{_0}}{3}\right)^2 p_{r_0}\alpha,\label{sigma4_exp}\\
&&\hspace{-0.4cm} p_{r_4} = -(p_{r_0}+\rho_{_0})\Psi_{_4}-\frac{1}{2}(p_{r_2}+\rho_{_2})\Psi_{_2} + \frac{1}{2}\sigma_{_4}.\label{p4_exp}
\end{eqnarray}
Note that all coefficients shown in Eqs. \eqref{phi2_exp}-\eqref{p4_exp} are consistent with those obtained for the isotropic case. Since when $\alpha=0$, Eqs. \eqref{phi2_exp}-\eqref{phi4_exp} and \eqref{p4_exp} reproduce the equalities reported in Ref. \cite{sotani2001} and with the equalities \eqref{sigma2_exp} and \eqref{sigma4_exp} being identically null. Moreover, by evaluating the perturbation equation to the second-order terms $\left[{\tilde H}_1''(0), {\tilde K}''(0), {\tilde W}''(0), {\tilde X}''(0)\right]$, we find that the following relationships must hold:
%
%\begin{widetext}
\begin{eqnarray}
&&\hspace{-0.4cm}\frac{\ell+3}{2}{\tilde H}_1''(0) - {\tilde K}''(0) + 8\pi(p_{r_0}+\rho_{_0})\frac{\ell+3}{\ell(\ell+1)}{\tilde W}''(0)= 4\pi\left[\frac{1}{3}(2\ell+3)\rho_{_0}- p_{r_0}\right]{\tilde H}_1(0)\nonumber\\
&&\hspace{-0.4cm}+ 8\pi(p_{r_0}+\rho_{_0})Q_1 + \frac{1}{2}Q_0- \frac{8\pi}{\ell}\left(p_{r_2}+\rho_{_2}\right){\tilde W}(0),\\
&&\hspace{-0.4cm}\frac{\ell+2}{2}{\tilde K}''(0)- \frac{\ell(\ell+1)}{4}{\tilde H}_1''(0) - 4\pi(p_{r_0}+\rho_{_0}){\tilde W}''(0)= 4\pi\left[p_{r_2}+\rho_{_2} + \frac{8\pi}{3}\rho_0(p_{r_0}+\rho_{_0})\right]{\tilde W}(0)\nonumber \\
&&\hspace{-0.4cm}+\frac{1}{2}Q_0+ \left(\frac{4\pi}{3}\rho_{_0} + 4\pi p_{r_0}\right){\tilde K}(0),\label{B.21}\\
%%%%%%%%%%%%%%%%%%%%%%%%%%%%%%%%%
&&\hspace{-0.4cm}\frac{\ell+2}{2}\tilde{X}''(0) + \frac{\ell(\ell+1)}{8}e^{\Psi_{_0}}(p_{r_0}+\rho_{_0})\tilde{H}_1''(0) + e^{\Psi_{_0}}(p_{r_0}+\rho_{_0})\left[\frac{\omega^2}{2}e^{-2\Psi_{_0}}+2\pi(p_{r_0}+\rho_{_0})\right.\nonumber \\
&&\hspace{-0.4cm}\left.-\frac{\ell+2}{2}\Psi_{_2}\right] \tilde{W}''(0)=  \frac{\ell}{2}\left(\Psi_{_2}+\frac{p_{r_2}+\rho_{_2}}{p_{r_0}+\rho_{_0}}\right)\tilde{X}(0)+ e^{\Psi_{_0}}(p_{r_0}+\rho_{_0})\biggl[-\frac{\ell(\ell+1)}{2}\Psi_{_2}Q_1-\frac{1}{4}Q_0 \nonumber \\
&&\hspace{-0.4cm}- \left(\frac{\omega^2}{2}e^{-2\Psi_{_0}}\right)\tilde{H}_1(0)-\Psi_{_2}\tilde{K}(0)+ \biggl\{(\ell+2)\Psi_{_4} -2\pi(p_{r_2}+\rho_{_2})-\frac{16\pi^2}{3}\rho_{_0}(p_{r_0}+\rho_{_0})-\frac{4\pi}{3}\rho_{_0}\Psi_{_2} \nonumber \\
&&\hspace{-0.4cm}+\omega^2e^{-2\Psi_{_0}}\left(\Psi_{_2}-\frac{4\pi}{3}\rho_{_0}\right)-\frac{\sigma_4}{p_{r_0}+\rho_{_0}}\biggr\}\tilde{W}(0)\biggr],\\
%%%%%%%%%%%%%%%%%%%%%%%%%%%%%%%%%%%%%%
&&\hspace{-0.4cm}-\frac{1}{4}(p_{r_0}+\rho_{_0}){\tilde K}''(0) -\frac{1}{2}\left[p_{r_2}+\frac{\ell+3}{\ell(\ell+1)}\omega^2e^{-2\Psi_{_0}}\left(p_{r_0}+\rho_{_0}\right)\right]{\tilde W}''(0) - \frac{1}{2}e^{-\Psi_{_0}}{\tilde X}''(0)\nonumber\\
&&=-\frac{1}{2}e^{-\Psi_{_0}}\Psi_{_2}{\tilde X}(0) + \frac{1}{4}(p_{r_2}+\rho_{_2}){\tilde K}(0)+ \frac{1}{4}\left(p_{r_0}+\rho_{_0}\right)Q_0+ \left[\frac{\omega^2}{\ell}e^{-2\Psi_{_0}}\left[\frac{1}{2}(p_{r_2}+\rho_{_2})\right.\right.\nonumber\\
&&\left.\left.- (p_{r_0}+\rho_{_0})\Psi_{_2}\right]+p_{r_4} - \frac{4\pi}{3}\rho_{_0}p_{r_2}\right]{\tilde W}(0)- \frac{1}{2}\omega^2e^{-2\Psi_{_0}}(p_{r_0}+\rho_{_0})Q_1,
\end{eqnarray}
%\end{widetext}
with $Q_0$ and $Q_1$ being defined by
\begin{eqnarray}
&&\hspace{-0.4cm}Q_0=\frac{4}{(\ell+2)(\ell-1)}\biggl[-8\pi e^{-\Psi_{_0}}{\tilde X}(0) - \left(\omega^2e^{-2\Psi_{_0}}+\frac{8\pi}{3}\rho_{_0}\right){\tilde K}(0)+ \left[\omega^2e^{-2\Psi_{_0}}\right.\nonumber \\
&&\hspace{-0.4cm}\left.-\ell(\ell+1)\left(\frac{2\pi}{3}\rho_{_0}+2\pi p_{r_0}\right)\right]{\tilde H}_1{(0)}\biggr],\\
&&\hspace{-0.4cm}Q_1 = \frac{2}{\ell(\ell+1)}\left[\frac{4\pi}{3}(\ell+1)\rho_{_0}{\tilde W}(0) - \frac{3}{2}{\tilde K}(0)+ \frac{e^{-\Psi_{_0}}}{\Gamma p_{r_0}}{\tilde X}(0)\right].
\end{eqnarray}
It is important to mention that all equations presented in this section agree with those presented in \cite{detweiler1983,sotani2001} for the isotropic case. In addition, the expansion of Eq. \eqref{B.21} to the leading order is the same as Eq. \eqref{B.6}.

\end{document}